\documentclass[a4paper,11pt]{article}

\usepackage{jheppub} 

\usepackage[T1]{fontenc} 

\usepackage{caption,subcaption}
\usepackage{tikz}
\usetikzlibrary{calc}
\usetikzlibrary{arrows,decorations.pathmorphing,patterns}
\usepackage{feynmp}
\usepackage{soul}
\usepackage{tikz-feynman}
\tikzfeynmanset{compat=1.0.0}
\numberwithin{equation}{section}
\usepackage{version}
\usepackage{mathrsfs} 

\usepackage{hyperref}
\hypersetup{
	colorlinks=true,
	linkcolor=blue,
	filecolor=magenta,      
	urlcolor=blue,
	citecolor=magenta
}
\def\pr{\partial}

 \csname
@addtoreset\endcsname{equation}{section}

\def\a{\alpha} 
  
\def\g{\gamma} 
\def\G{\Gamma}
\def\d{\delta} 
\def\D{\Delta}
\def\e{\epsilon} 
\def\ve{\varepsilon}

\def\th{\theta} 
\def\Th{\Theta}

\def\l{\lambda}

\def\m{\mu}
\def\n{\nu}

\def\r{\rho}

\def\S{\Sigma}

\def\f{\phi}

\def\vf{\varphi}
\def\O{\Omega}
\def\o{\omega}

\def\cA{{\cal A}}
\def\cB{{\cal B}}

\def\cD{{\cal D}}

\def\cI{{\cal I}}
\def\cJ{{\cal J}}

\def\cL{{\cal L}}

\def\cO{{\cal O}}

\def\cR{{\cal R}}

\def\cT{{\cal T}}

\def\bz{{\bar{z}}}

\def\tA{{\tilde{A}}}

\def\hA{{\hat{A}}}
\def\ha{{\hat{a}}}

\def\te{{\tilde{\e}}}

\def\he{ \hat \e}
\def\hl{ \hat \l}

\def\dg{\sqrt{-g}\,}

\def\lp{\l^{\prime}}
\def\p{\prime}

\title{\boldmath ${\mathcal{O}(r^N)} $ two-form asymptotic symmetries and renormalized charges}%
\author[a]{Matteo Romoli}

\affiliation[a]{Roma Tre University and INFN Sezione di Roma Tre, via della Vasca Navale, 84 I-00146 Roma, Italy}

\emailAdd{matteo.romoli@uniroma3.it}

\abstract{We investigate $ \mathcal{O}\left( r^N \right) $ asymptotic symmetries for a two-form gauge field in four-dimensional Minkowski spacetime. By employing symplectic renormalization, we identify $ N $ independent asymptotic charges, with each charge being parametrised by an arbitrary function of the angular variables. Working in Lorenz gauge, the gauge parameters require a radial expansion involving logarithmic (subleading) terms to ensure nontrivial angular dependence at leading order. At the same time, we adopt a setup where the field strength admits a power expansion, allowing logarithms in the gauge field expansions within pure gauge sectors. The same setup is studied for electromagnetism.}

\notoc

\begin{document}
	
	\maketitle
	\newpage
	\tableofcontents
	\section{Introduction}
	Over the past decade, there has been a renewed interest in the subject of asymptotic symmetries in gravity and gauge theories, mostly due to the established connection with soft theorems and memory effects (see \cite{Strominger:2017zoo} for a review), elegantly described by the infrared triangle. In fact, the concept of asymptotic symmetry has been extended to other theories, including electrodynamics \cite{He:2014cra,Pasterski:2015zua}, Yang-Mills theory \cite{Strominger:2013lka,He:2015zea}, $ p $-forms theories \cite{Campiglia:2018see,Francia:2018jtb, Afshar:2018apx} and higher-spin theories \cite{Campoleoni:2017mbt, Campoleoni:2018uib}. It has also been generalised to arbitrary spacetime dimension \cite{Kapec:2015ena,He:2019jjk,He:2019pll,Henneaux:2019yqq,Campoleoni:2019ptc,Campoleoni:2020ejn,Fuentealba:2023huv}, different backgrounds \cite{Esmaeili:2019mbw,Esmaeili:2021szb,Campoleoni:2023eqp} and to more general contexts \cite{Barnich:2010eb,Campiglia:2017dpg,Agriela:2023dnw,Manzoni:2024agc}. 
	
	Building on the ideas that led to the formulation of superrotations \cite{Barnich:2009se,Barnich:2010ojg}, asymptotic symmetries have been generalised to higher orders, where we define the order of an asymptotic symmetry in relation to the radial power of the corresponding asymptotic parameter in Cartesian components. For instance, supertranslations are $ \cO(1) $ (or $ \cO(r^{0}) $) asymptotic symmetries, while superrotations are $ \cO(r) $. In this respect, the idea at the basis of the infrared triangle is that the semiclassical Ward Identity stemming from an $ \cO(r^{n}) $ asymptotic symmetry is equivalent to the sub$ {^n} $-leading soft theorem. We can find analyses of $ \cO(r) $ asymptotic symmetries in electromagnetism \cite{Campiglia:2016hvg}, recently generalised to arbitrary $ \cO(r^{N}) $ in \cite{Peraza:2023ivy}. The non-abelian case was studied at $ \cO(r) $ in \cite{Campiglia:2021oqz}, and recently extended to arbitrary higher order in \cite{Nagy:2024dme,Nagy:2024jua}. For gravity, $ \cO(r^2) $ asymptotic symmetries were studied in \cite{Campiglia:2016efb}, exploring the connection with sub-subleading soft graviton theorem. In this work, we present the $ \cO(r^{N}) $ generalisation of two-form asymptotic symmetries, whose $ \cO(1) $ formulation was studied in the context of the scalar duality in \cite{Campiglia:2018see, Francia:2018jtb} and recently refined in \cite{Ferrero:2024eva}.  
	
	In its most basic incarnation, the program of asymptotic symmetries consists in finding, among the residual gauge symmetries, the ones that preserve the boundary conditions, {\it{i.e.}} the chosen (field) falloffs, and still act nontrivially on asymptotic field configurations. Since we are interested in gauge fields, we discuss boundary conditions at null infinity. In particular, we focus our analysis on $ \cI^+ $, which is the region characterized by the limit $ r\to \infty $ at fixed retarded time $ u = t-r $. The falloffs are determined by asking finiteness of physical quantities as well as the presence of all physically reasonable solutions. For instance, one of the conditions considered in electromagnetism is that the long-range electric field falls off as $ \tfrac{1}{r^{2}} $, while for gravity the characterization of the boundary is captured by the concept of asymptotic flatness at null infinity \cite{Bondi:1962px,Sachs_Symmetries,Penrose:1962ij,Penrose:1964ge,Penrose:1965am}. Typically, one starts by determining the falloffs of the field strength of the theory under investigation and then deduces the gauge fields falloffs. This operation is not entirely unambiguous, as field falloffs can be always generalised including potentially overleading pure gauge terms. This flexibility generally allows for arbitrary extension of field falloffs and, consequently, the generalisation of the order of the asymptotic parameter that preserves these falloffs\footnote{This is not the case, for instance, of radial gauge in electromagnetism, where the gauge condition is ``too strong'' to allow for $ \cO(r^N) \; (N>0)$ asymptotic parameters.}. 
	
	Another subject of interest, related to the previous discussion, regards the choice of the expansion for fields and parameters. Early works on this topic considered a $ \tfrac{1}{r} $ expansion, but more recent works have included logarithmic terms \cite{Hirai:2018ijc,He:2019jjk,He:2019pll,Campoleoni:2019ptc}, which seem unavoidable in certain setups, such as when working in Lorenz gauge. The Lorenz gauge, which we adopt, is indeed peculiar: in order to find asymptotic symmetries, the parameter must have logarithmic terms in its expansion, while this is not necessary for what concerns the field components. In favour of simplicity, one might be tempted to admit two different expansions for field components and parameters, considering logarithmic terms only in the latter, as it is done in \cite{Campoleoni:2019ptc}. In that case, however, it is not possible to preserve the falloffs without losing asymptotic symmetries. For this reason, we work with fields and parameters admitting the same expansion but, as we comment in detail below, the logarithms in the field components appear only in pure gauge sectors, so that the field strength admits a simple $ \tfrac{1}{r} $ expansion. Details are added in \autoref{sec:em} and \autoref{sec:2f}.
	
	When dealing with higher order asymptotic symmetries, however, a relevant issue occurs: the asymptotic charge turns out to be divergent. These divergences can be understood as ambiguities of the presymplectic potential and therefore renormalized in the spirit of \cite{Freidel:2019ohg}. We apply this procedure to our asymptotic charges, aiming to cancel all the divergences while leaving unchanged the finite parts. 
	
	There are several motivations for this work. To begin with, $ \cO(r^N) $ two-form asymptotic symmetries are interesting because of the existence of the scalar duality, since in $ D=4 $ a two-form is dual (on-shell) to a scalar field. Hence, understanding these higher order asymptotic symmetries could be the key to understand asymptotic symmetries for scalars. And since scalar theories represent the simplest models in physics, these results could help in a deeper understanding of the connection between asymptotic symmetries and physical effects. 
	
	Furthermore, it has been recently pointed out in \cite{Ferrero:2024eva} how $ \cO(1) $ asymptotic symmetries of electromagnetism are mapped, via the double copy, to BMS supertranslations and $ \cO(1) $ two-form asymptotic symmetries. One might wonder whether this result is also generalisable, for instance, to $ \cO(r) $, giving the two-form counterpart of superrotations. However, as mentioned, two-form asymptotic symmetries have been analysed only at $ \cO(1) $. Other double-copy perspectives on asymptotic symmetries can be found in \cite{Campiglia:2021srh,Nagy:2022xxs} with focus on the self-dual sector.
	
	Additionally, let us observe that the two-form provides the simplest example of a theory possessing a gauge-for-gauge redundancy, namely the gauge parameter has itself a gauge symmetry. We thus might expect some peculiarities that are not present in the electromagnetic case. From our analysis, however, it seems that no particular role is played by the gauge-for-gauge parameter, except that of simplifying some computations.
	
We begin our analysis with the Maxwell field and then extend it to the two-form theory. Although much of the analysis regarding the spin-one case can be understood as a review of \cite{Peraza:2023ivy}, our setup is slightly different and the symplectic renormalization require some extra comments. We then explore the two-form case, first reviewing the existing results in the literature about $ \cO(1) $ asymptotic symmetries  \cite{Campiglia:2018see,Francia:2018jtb,Ferrero:2024eva} and then generalising the discussion to arbitrary $ \cO(r^N) $. The symplectic renormalization is presented in \autoref{sec:ren}.
	\section{Preliminaries}
	\subsection{Notation and conventions}
	We focus on $ D=4 $ and adopt the mostly plus convention for the metric. For the discussion of asymptotic parameter, we employ retarded Bondi coordinates $(u, r, z^i)$, with $r$ being the radial coordinate, $u = t-r$ the retarded time and $z^i\; (i=1,2)$ the angular coordinates, which we parameterise by means of stereographic projection in terms of $ (z,\bz) $, related to the standard angles as
	\begin{equation}
		z = e^{i\f} \cot\frac{\th}{2}, \qquad \bar{z} = e^{-i\f} \cot\frac{\th}{2}\, ,  
	\end{equation}
	with $\theta \in [0, \pi]$ and $\phi \in [0, 2 \pi)$.
	
	The Minkowski metric reads 
	\begin{equation}\label{metric}
		ds^2=-du^2-2du\, dr + r^2 \gamma_{ij}dz^i\,dz^j,
	\end{equation} 
	where $\gamma_{ij} $ is the unit metric on the two-sphere, which is anti-diagonal in terms of $(z,\bz)$, with 
	\begin{equation}
		\gamma_{z\bar{z}} = \frac{2}{(1+z\bar{z})^2}.
	\end{equation}

	The covariant derivative with respect to $\gamma_{ij} $ is denoted with $D_i$ and the two-sphere Laplace operator is $ \D = D_i D^i $. At the same time, we use $ \nabla_\m $ for the covariant derivative respect to the metric \eqref{metric} and $ \Box = \nabla_\m\nabla^\m $ for the d'Alembert operator. Divergences are often denoted as $ D\cdot V := D^i V_i $ and $ \nabla\cdot V := \nabla^\m V_\m$.
	
	The non-vanishing Christoffel symbols are 
	\begin{equation}
		\G^{u}_{ij} = -	\G^{r}_{ij} = r\g_{ij},\qquad \G^{i}_{rj} = \frac{1}{r}\d^i_j,\qquad \G^z_{zz}= {\bar{\G}^{\bz}_{\bz\bz}} =-\frac{2z}{1+z\bz} \,.
	\end{equation} 

We often make use of the following commutation relation, which we write here explicitly:
\begin{equation}
	D_i \D V_j =  (\D -3) D_i V_j+ 2 \g_{ij} D_k V^k.
\end{equation}

Future null infinity $ \cI^+ $ is reached by sending $ r\to \infty $ (or $ t\to\infty $) with fixed $ u $. It is a null hypersurface with $ \mathbb{R}\times S^2 $ topology. Sending $ u\to\pm\infty $ we reach $ \cI^+_{\pm} $. Fields and parameter are expanded near $ \cI^+ $. The most general radial expansion we write is 
\begin{equation}\label{gen_expansion}
	\vf = \sum_{n_1} \frac{\vf^{(n_1)}}{r^{n_1}}+\sum_{n_2} {\hat \vf}^{(n_2)} \frac{\log r}{r^{n_2}},
\end{equation}
where the coefficients of the radial expansion are always denoted with a superscript ``$(n)$'' and, in particular, we use the hat for the logarithmic coefficients. 

When discussing charges and renormalization we employ $ (t,u,z,\bz) $ coordinates, since we are interested in the large $ t- $behaviour around $ \cI^+ $. The expansion coefficients are denoted with a subscript ``$ (n) $'' and
\begin{equation}\label{t_expansion}
	\vf = \sum_{n_1} {t^{n_1}{\vf_{(n_1)}}}+{\log t}\sum_{n_2}  {t^{n_2}}\,{\hat \vf}_{(n_2)}.
\end{equation}
	
To describe the behaviour near $ \cI^+_- $ we consider only power expansions that we denote as
\begin{equation}
	\vf^{(n)} = \sum_{m} u^{m} \vf^{(n,m)}.
\end{equation}
In some cases this expansion is valid for all values of $ u $ and not only for $ u\to-\infty $,  but we use the same notation, always specifying its meaning. 
	
\subsection{Covariant phase space formalism}
	The most important tool in order to understand whether an asymptotic symmetry has a physical interpretation is the asymptotic charge. We introduce
	this notion employing the covariant phase space formalism, whose main idea is to combine the calculus in both spacetime and field space. This technique was introduced in \cite{GAWEDZKI1972307,Kijowski:1973gi,Kijowski:1976ze} and later refined in \cite{Lee:1990nz,Wald:1993nt,Wald:1999wa}. Discussions with a particular focus on asymptotic symmetries can be found in \cite{Barnich:2001jy,Avery:2015rga}. Reviews can be found, for instance, in \cite{Compere:2018aar,Ruzziconi:2019pzd,Ciambelli:2022vot}.
	
	On a differentiable manifold $ M $, the space of forms induce the de Rham cohomology, where we denote with $ d $ and $ i $ the exterior derivative and interior product on this complex,
	respectively. The exterior derivative is assumed to be nilpotent, i.e.\ $ d^2 = 0 $. The Lie derivative along $ \xi \in TM $ is given by
	\begin{equation}
		\cL_\xi = d\,i_\xi + i_\xi d.
	\end{equation}
	
	These notions can be introduced also in the space of all possible field configurations $ \G $. In this case, we denote with $ \d $ the exterior derivative, assumed to be nilpotent as well, while we use $ I_{V} $ for the interior product, so that the Lie derivative reads 
	\begin{equation}
		\mathfrak{L}_V := \d \,I_V + I_V\d\,.
	\end{equation}
	
	Given a Lagrangian theory, whose action reads 
	\begin{equation}
		S = \int_M L,
	\end{equation}
	we can write the way $ L $ transforms under a generic field variation $ \vf \to \vf + \d \vf $ as
	\begin{equation}\label{variation_L}
		\d L= \text{EOM} \d \vf + d\th,
	\end{equation}
	with $ \th $ being the presymplectic potential. We define the presymplectic (field-space) two-form as 
	\begin{equation}
		\o = \d \th,
	\end{equation} 
	and the symplectic two-form
	\begin{equation}
		\O = \int_\S \o,
	\end{equation}
	with $ \S $ being an arbitrary Cauchy surface. 
	
	Assuming trivial cohomology in the space of $ 1 $-forms on $ \G $, a vector $ V \in T\G $ is called a symplectomorphism or Hamiltonian vector field if $ \mathfrak{L}_V \o = 0 $. For a gauge symmetry, we can define the current as 
	\begin{equation}
		J_V = I_V \th
	\end{equation}
	and the charge 
	\begin{equation}\label{charge}
		Q = \int_\S J_V. 
	\end{equation}
	Furthermore, for a gauge symmetry the current is in general an on-shell total derivative and the charge can be written as a integral over $ \pr \S $ by means of the Stokes theorem.
	
	 The presymplectic potential admits two types of ambiguities
	\begin{equation}\label{ambig}
		\th \to \th + \d \Xi + d\Upsilon.
	\end{equation}
	The first does not change $ \o $, since $ \d $ is nilpotent, and corresponds to the addition of a boundary term to the Lagrangian $ L\to L +d\Xi $. The second modifies $ \o\to \o + d\d \Upsilon $  but does not affect $ d\th $ in \eqref{variation_L}, since $ d $ is nilpotent.
	
The existence of these ambiguities is fundamental in order to discuss asymptotic charges. In fact, as we discuss later, some divergences occur when dealing with higher order asymptotic symmetries. Therefore, it is not possible to perform the limit to define asymptotic charges. However, we can absorb the divergent terms in ambiguities of the presymplectic potential, while leaving the finite and non-vanishing parts of the asymptotic charges unchanged; the latter provide the physical content of the asymptotic symmetries.
	
	\subsection{Asymptotic charges}
	\paragraph{Electromagnetism.}
	Considering the Lagrangian
	\begin{equation}
		L = -\frac{1}{4}F_{\m\n}F^{\m\n}\,,
	\end{equation}
	the presymplectic potential current reads 
	\begin{equation}
		\th_{e}^{\m} = -\dg F^{\m\n} \d \cA_\n.
	\end{equation}
	Given a gauge transformation $ \d \cA_\n = \pr_\n \e  $ the current is an on-shell total derivative
	\begin{equation}
		\cJ_e^\m = -\pr_\n(\dg F^{\m\n}\e ) \,.
	\end{equation}
	We consider a surface $ \S_t $ at constant $ t = u+r $ and then study the limit $ t\to \infty $. This defines the asymptotic charge, evaluated in $ (u,t,z,\bz)-$coordinates, which is
	\begin{equation}\label{Q_e}
		\begin{split}
			Q_e =& \lim\limits_{t\to\infty}-\int_{\S_t} du dz d\bz \, \pr_\m ( \dg F^{t\m}\e)\\ =& \lim\limits_{t\to\infty} -\int_{\S_t} du dz d\bz\g_{z\bz} (\pr_u-\pr_r) ( r^2 F_{ur}\e),
		\end{split}
	\end{equation}
	where in the last step we substituted the expression in terms of Bondi components and neglected a total two-sphere divergence.  
	
	At the same time, we can write a dual asymptotic magnetic charge, which is 
	\begin{equation}\label{Q_m}
		{\tilde Q}_m =  \int_{\S_t} du dz d\bz (\pr_u-\pr_r) ( \dg {\tilde F}_{ur}\te)\,,
	\end{equation}
	where $ \tilde F $ is the dual field strength, defined as through the Hodge dual $ \tilde F = \star F $, and $ \te $ is the gauge parameter of the dual gauge field $ \tilde A $, with $ \tilde F = d \tA  $. 
	
	 However, it should be noted that the definition \eqref{Q_m} is not entirely correct. As studied in detail in \cite{Freidel:2018fsk}, the charge \eqref{Q_m} is not derived from a well-defined canonical analysis. To obtain its correct form, one should extend the covariant phase space by means of edge modes degrees of freedom. In particular, the correct magnetic charge is different from the one \eqref{Q_m} when the dual gauge parameter has singularities. 
	 
	We do not perform such an analysis here, as our primary interest lies in the two-form and the comparison with the electric charges. The correct form of the charge is used in \cite{Peraza:2023ivy} for a $ \cO(r^n) $ dual gauge parameter. We leave this analysis for future investigations.
	
	\paragraph{2-form. }
	The Lagrangian is
	\begin{equation}
		L = -\frac{1}{6}H_{\m\n\r}H^{\m\n\r} \, \text{Vol}_M\,,
	\end{equation}
	so that the presymplectic potential reads 
	\begin{equation}
		\th_{B}^{\m} = -\dg H^{\m\n\r} \d \cB_{\n\r}\,.
	\end{equation}
	Given a gauge transformation $ \d \cB_{\n\r} = \frac{1}{2}(\pr_\n \l_\r-\pr_\r \l_\n) $ the current takes the form
	\begin{equation}
		\cJ^\m_B = -\pr_\n(\dg H^{\m\n\r}\l_\r ) . 
	\end{equation}
	Considering again a constant time surface $ \S_t $ , the asymptotic charge is defined as the limit $ t\to\infty $ of the quantity
	\begin{equation}\label{B_charge}
	Q_{B}= \int_{\S_t} du dz d\bz \g_{z\bz}(\pr_u-\pr_r)\left (r^2 H^{uri}\l_i\right)
	\end{equation}
	where we are neglecting again a total two-sphere divergence.

	For higher-order asymptotic symmetries, there are $ t- $divergences which imply that the both $ Q_e $ and $ Q_{B} $ are not well-defined in the limit. However, we can employ symplectic renormalization in order to neglect these divergences and perform the limit.

\section{Electromagnetism}\label{sec:em}
	We start by deriving $ \cO(r^N) $ asymptotic symmetries for electromagnetism. Unlike \cite{Campiglia:2016hvg,Peraza:2023ivy}, which provide a similar setup to ours, we do not consider sources in our analysis and focus only on the soft part of the charges, with the aim of comparing the spin-one and two-form results. We motivate this choice noting how the double-copy results \cite{Ferrero:2024eva}, that partially inspired this work, are based on a dictionary that treats differently fields and sources \cite{Anastasiou:2014qba}. In addition to this, we notice that considering the source of \cite{Campiglia:2016hvg,Peraza:2023ivy}, namely the one of scalar electrodynamics  
	   \begin{equation}
		j_\m = ie\phi \cD_\m \phi^* + c.c
	\end{equation}
with $  \cD_\m \phi = \pr_\m \phi -ie\cA_\m \phi $, and assuming the standard behaviours near $ \cI^+ $, namely $ \phi =\cO(\tfrac{1}{r}) $ and $ \cA_u =\cO(\tfrac{1}{r}) $, we find a constraint on the $ u- $dependence of $ \f^{(1)} $, which comes from the condition $ j_u^{(2)}=0 $. 

Furthermore, we cannot treat $A_\m$ and $j_\m$ independently in an order-by-order analysis of the equations of motion, crucial to determine the correct falloffs. In this respect, we remark how some leading components of the current $ j_\m $ already involve $A_\m$, making it nontrivial to assume that the results align with those derived from the free equations of motion. For instance, in the case of $ \f^3 $ theory in $ D=4 $, as noted in \cite{Campiglia:2017xkp}, the standard falloff of scalar field $ \f = \cO( \tfrac{1}{r}) $,  is not consistent with the equations of motion in the presence of the interaction, while being fully compatible with the vacuum equations of motion. 

A full analysis is left for future investigation, while for the moment we restrict our study to the scenario where all quantities satisfy the vacuum equations of motion, analysed in \autoref{app:eom}. Let us observe that this setup applies also to the case of solutions far from localised sources, to which the vacuum equations of motion provide a good approximation \cite{Cristofoli:2021vyo}.

The falloffs analysis starts by asking compatibility with some physical requests about the field strength:
	\begin{equation}\label{phys}
		F_{ur} = \cO(r^{-2}), \qquad F_{ui} = \cO(1), \qquad F_{ri} = \cO(r^{-2}), \qquad F_{ij} = \cO(1).
	\end{equation}
	Moreover, we propose a key ansatz regards the expansion of the field strength components:
	\begin{equation}\label{F_exp}
		F_{\m\n} = \sum_n \frac{F_{\m\n}^{(n)}(u,z,\bz)}{r^n},
	\end{equation} 
	where we add details on the $ u- $dependence in the next subsection. Hence, logarithms are not included in the field strength expansion, unlike the gauge parameter and the gauge field components. This choice is motivated by observing how logarithms in \eqref{F_exp} are unnecessary. In fact, they are required in the gauge parameter because otherwise the Lorenz gauge condition does not allow for a non-trivial $ \cO(1) $ term. Consequently, logarithms are added in the field components or otherwise the falloffs cannot be preserved (see, for instance, section 5 of \cite{Campoleoni:2019ptc}). However, this issue is specific to the Lorenz gauge but does not arise, for instance, in retarded radial gauge or in radiation gauge. In these well-studied and somehow simpler cases, the parameter does not require logarithms, therefore there is no need to consider these terms in the gauge field components and, accordingly, no reason to add them in the field strength. And since the field strength is a gauge invariant quantity, we do not expect the expansion considered to be gauge-dependent. This choice is one of the main difference with \cite{Peraza:2023ivy}.
	
	We work in Lorenz gauge
	\begin{equation}
		\nabla^\m \cA_\m = 0,
	\end{equation}
	so that the residual gauge parameter $ \e $ obeys to
	\begin{equation}
		\Box \e=0
	\end{equation}
	and the Maxwell equations reduces to 
	\begin{equation}
		\Box \cA_\m=0. 
	\end{equation}

	\subsection{Falloffs}
	The choice \eqref{F_exp} has a crucial consequence on the logarithmic terms of the gauge field components: they are pure gauge terms. Therefore, the gauge field admits the general expansion
	\begin{equation}\label{A_expansion}
		\cA_\m =  A_{\m}+\pr_\m a =\sum_{n,n_1,n_2} \left [\frac{A_\m^{(n)}}{r^{n}}+ \pr_\m\left(\frac{a^{(n_1)}}{r^{n_1}} + \ha^{(n_2)}\frac{\log r}{r^{n_2}}\right)\right ],
	\end{equation}
	where we factored out all the pure gauge sector. The Lorenz gauge condition implies that the $ a $-series satisfies the wave equation.
	
	 Only the first series in \eqref{A_expansion} contributes to the field strength, allowing us to determine the falloffs of $ A_\m $ from \eqref{phys}, which are :
	\begin{equation}\label{A_phys_falloffs}
		A_u = \cO(r^{-1}),\qquad A_r = \cO(r^{-2}) \qquad A_i = \cO(1),
	\end{equation}
	where constant terms are neglected\footnote{See \autoref{subs:glob} for further comments.}. It is straightforward to verify that these falloffs are consistent with the equations of motion. In fact, since the logarithms appear in pure gauge sectors, it is sufficient to study the equations of motion for a $ \tfrac{1}{r} $ expansion.  Specifically, $ A_i $ has the role of the free Cauchy data, as it remains unconstrained by the equations of motion, meaning that $ A_i^{(0)}=A_i^{(0)}(u,z,\bz) $ is arbitrary on $ \cI^{+} $. We assume that, as we approach $ \cI^{+}_{-} $,  $ A_i^{(0)} $ tends to a well-defined function of $ (z,\bz) $ as
	\begin{equation}\label{free_scri}
		 A_i^{(0)}(u,z,\bz)  = \sum_{m=-\infty}^{0}u^{m}A_i^{(0,m)}(z,\bz).
	\end{equation}
	Although other expansions\footnote{\label{foot:tree}In \cite{Peraza:2023ivy} it is assumed that the term $ A_i^{(0)} $ satisfies the so-called tree-level assumption, namely
		\begin{equation}\label{tree_ass}
			A_i^{(0)} = A_i^{(0,0)}+\cO(|u|^{-\infty})
		\end{equation}
		in the limit $ u\to\pm \infty $, meaning that 
		$ \lim\limits_{u\to\pm\infty} u^{n}\pr_u A_i =0\;\; \forall\, n>0  $. This assumption is not required for the purposes of this work and therefore we consider a power expansion. See the conclusions for further comments on other possible choices.} are possible, the structure in \eqref{free_scri} simplifies the discussion of renormalization.  In particular, this assumption has consequences on the expansion of the presymplectic potential in \eqref{th-uexp} and the consequent renormalization procedure.
	
	We can also rearrange the expansion \eqref{A_expansion} as
	\begin{equation}
		\cA_\m = \sum_{n} \frac{\cA_\m^{(n)} }{r^n}+\sum_{n'} {\hat{\cA}}_\m^{(n')} \frac{\log r}{r^{n'}},
	\end{equation}
	where it is important to remark that the second series is entirely pure gauge, while the first series receives contribution from both $ A_\m$ and $ a $.
	
By allowing for overleading pure gauge terms, we are extending the space of asymptotic field configuration $ \G $ and therefore we are generalising the presymplectic potential, since the variation $ \d \cA_\m $ is within the space $ \G $. This turns out to be crucial for the discussion of charge renormalization.

\subsection{Field strength}
As shown in \eqref{Q_e}, the component $ F_{ur} = \cO(r^{-2}) $ enters in the asymptotic electric charge definition. Specifically, the assumption 
\begin{equation}
	F_{ur} = \sum_{n=2}^{\infty}\frac{F_{ur}^{(n)} }{r^{n}}
\end{equation}
implies that only the first series of \eqref{A_expansion} contributes to it. Using the equations of motion, we find
\begin{equation}\label{fur}
	\begin{split}
		F_{ur}^{(2)} &= D\cdot A^{(0)}\\
			F_{ur}^{(n)} &= -\frac{\D}{n-2} A_r^{(n-1)}-D\cdot A^{(n-2)}\qquad (n> 2),
		\end{split}
	\end{equation}
as well as
\begin{equation}\label{dfur}
	\begin{split}
		\pr_u F_{ur}^{(2)} &= \pr_u D\cdot A^{(0)} \\
		\pr_u F_{ur}^{(3)} &= -\frac{1}{2}\D D\cdot A^{(0)} \\
		\pr_u F_{ur}^{(n)} &=  \frac{\D}{2}\left(\frac{\D }{(n-2)(n-3)}+1  \right)A_r^{(n-2)}+\left(\frac{\D +(n-2)(n-3)}{2(n-2)}\right)D\cdot A^{(n-3)} \qquad (n>3). 
	\end{split}
\end{equation}
We can perform a further gauge fixing and express $ F_{ur}^{(n)} $ only in terms of $ D\cdot A $. This is discussed in the rest of the section and in detail in \autoref{app:eom_sol}.

These equations, together with the assumption \eqref{free_scri}, fixes the $ u- $dependence around $ u\to-\infty  $ to be 
\begin{equation}\label{fur_exp}
	F_{ur} = \sum_{n=2}^{\infty} \frac{1}{r^{n}} \left [\sum_{m=0}^{n-2} u^{m} F_{ur}^{(n,m)}(z,\bz) + f_{ur}^{(n)}(u,z,\bz)\right ]
\end{equation}
where $ f_{ur}^{(n)}(u,z,\bz) \to 0 $ for $ u\to-\infty $ and, consistently with \eqref{free_scri}, has a $ u- $power expansion. On $ \cI^{+}_+ $ we assume that the field strength goes to zero because we are not considering massive charged particles.

The dual tensor $ {\tilde F}_{ur}:=(\star F)_{ur}  $ is defined as
\begin{equation}
{\tilde F}_{ur}	= \ve_{urij}F^{ij} =\frac{\ve^{ij}}{r^2}F_{ij},
\end{equation}
where $ \ve^{ij} $ is the antisymmetric Levi-Civita tensor on the two-sphere. 

In order to analyse it, let us introduce a particularly useful splitting which generally valid for every two-sphere vector
\begin{equation}\label{split_A}
	A_i = D_i A + \ve_{ij} D^j A^{\prime}.
\end{equation}
This is the Helmholtz decomposition on the two-sphere which, in stereographic coordinates, takes the simple form
\begin{equation}
	\begin{split}
		&A_z = D_z (A +  A^{\prime}),\\
		&A_\bz = D_\bz (A -  A^{\prime}).
	\end{split}
\end{equation}
This splitting is particularly well-suited for analysing asymptotic symmetries since these two components decouple in the asymptotic charges. Indeed, the tensor $ F_{ur} $, that appear in the asymptotic electric charge definition, depends on $ D\cdot A = \D A $, while the dual tensor, which enters in the magnetic charge, depends exclusively on $ {\ve^{ij}}F_{ij} = 2\D A^{\prime} $. 

\subsection{Asymptotic symmetries}
\subsubsection{\boldmath{$ \cO(1) $} asymptotic symmetries}
Working in Lorenz gauge, the residual gauge parameter satisfies the wave equation
	\begin{equation}\label{scalar_wave}
		\Box \e = 0.
	\end{equation}
	In order to find a $ \cO(1) $ term with arbitrary dependence from angular variables, a standard power expansion is not sufficient\footnote{In fact, the wave equation for a parameter $ 	\e = \sum_{n=0}^{\infty} \frac{\e^{(n)}}{r^n} $ implies $ \D\e^{(0)}(z,\bz)=0 $, thus showing that only constants are $ \e^{(0)} $ are allowed. In $ D>4 $ we do not even find constants \cite{Campoleoni:2019ptc}.}. A solution is to use a polyhomogeneous expansion
	\begin{equation}\label{lorenz_par}
		\e = \sum_{n=0}^{\infty} \frac{\e^{(n)}}{r^n} + \sum_{n=1}^{\infty} \he^{(n)} \frac{\log r}{r^n} .
	\end{equation}

	This parameter does not preserve the falloffs \eqref{A_phys_falloffs}. We thus modify them to be
	\begin{equation}\label{1_falloff_lorenz}
		\cA_u = \cO(\tfrac{\log r}{r}), \qquad \cA_r = \cO(\tfrac{\log r}{r^2}),\qquad \cA_i = \cO(1),
	\end{equation}
	where the overleading terms respect to \eqref{A_phys_falloffs} are understood as pure gauge terms. Note that, if this were not the case, we would have found $ F_{ri} = \cO(\tfrac{\log r}{r^2}) $, overleading respect to \eqref{phys}, or should have imposed a constraint\footnote{To be precise, this constraint should be imposed only on the $ u-$independent part of $ \hat{\cA}_r^{(2)} $ and $ \hat{\cA}_i^{(1)} $, because for the $ u-$dependent part it comes from the equations of motion.}  on $ \hat{\cA}_r^{(2)} $ and $ \hat{\cA}_i^{(1)} $, namely $	\hat{\cA}_i^{(1)}  = -D_i \hat{\cA}_{r}^{(2)}$. According to the notation \eqref{A_expansion}, the falloffs \eqref{1_falloff_lorenz} consist in
	\begin{equation}
		\begin{split}
			&	A_u = \cO(r^{-1}),\qquad A_r = \cO(r^{-2}) \qquad A_i = \cO(1), \qquad a = \cO\left (1\right )\,,
		\end{split}
	\end{equation}  
	with $ a $ polyhomogeneously expanded as in \eqref{A_expansion}.
	
	The equation \eqref{scalar_wave} for a parameter of the form \eqref{lorenz_par} implies for the first orders that
	\begin{equation}\label{sol_lor_sc}
		\begin{split}
			& \e^{(0)}=\e_{c}^{(0)}(z,\bz),\\
			&\he^{(1)}= \frac{1}{2} u \D \e^{(0)} + \he_c^{(1)}(z,\bz),\\
			&\e^{(1)} = \e^{(1)}(u,z,\bz),
		\end{split}
	\end{equation}
	where the subscript ``$ c $ '' denotes a $u-$integration constant, that appears in general at every order, while $\e^{(1)}$ is completely free. All the functions that appear with a subscript ``$ c $ '' depend only on $ z $ and $ \bz $. In particular, the $ u- $dependent part of $\e^{(1)}$ can be used to fix $A_u^{(1)}=0$, leaving only $ \e^{(1)}_c(z,\bz) $ and all the $ \e^{(n)}_c(z,\bz) $ for $ n\geq1 $ can be used to gauge away $ u- $independent components of $ A_r^{(n)} $. In particular, we use the various $ \e^{(n)}_c $ to reach the condition 
		\begin{equation}
			A_r^{(n,0)}(z,\bz)=0,
		\end{equation} 
	at $ u\to-\infty $, meaning that we are cancelling the $ u- $independent part at $ \cI^+_- $. Comments are added in \autoref{app:eom_sol}. Having used all the subleading terms in $ \e $, we have reached a complete gauge fixing.
	
	 The parameter \eqref{sol_lor_sc} preserve the falloffs \eqref{1_falloff_lorenz} and, as we discuss below, it corresponds to the standard $ \cO(1) $ asymptotic charge.
	\subsubsection{\boldmath{$ \cO(r) $} asymptotic symmetries}
	We are now interested in a $ \cO(r) $ asymptotic parameter which, however, would violate the falloffs \eqref{1_falloff_lorenz}. Therefore, we consider	\begin{equation}\label{lorenz_falloff_2}
		\cA_r=\cO(1), \quad \cA_u=\cO(1), \quad \cA_i=\cO(r),
	\end{equation}
	where all the leading terms are now in pure gauge sectors, meaning that we consider $ a = \cO(r) $ respect to \eqref{A_expansion}. The parameter takes the form
	\begin{equation}\label{lorenz_par_2}
		\e = \sum_{n=-1}^{\infty} \frac{\e^{(n)}}{r^n} + \sum_{n=1}^{\infty} \he^{(n)} \frac{\log r}{r^n},
	\end{equation}
	where we emphasize that the logarithmic series starts again from $n=1$. In fact, the second series is necessary to find a $\cO(r^{0})$ term with arbitrary angular dependence, but then the $\cO(r)$ term can have it without requiring additional terms\footnote{We could, in principle, consider a $\log r$ term in the series \eqref{lorenz_par_2}, but in order for it to have a non-trivial angular dependence we should add a third series $$ \sum_{n=1}^{\infty} \tilde \e^{(n)} \frac{\log^{2} r}{r^{n}} $$.}
		
	For what concerns the gauge parameter, the wave equation implies
	\begin{equation}\label{o(r)_as}
		\begin{split}
			&\e^{(-1)} = \e_{c}^{(-1)} (z,\bz),\\
			&\e^{(0)} = u\frac{\D+2}{2}\e_{c}^{(-1)}+\e^{(0)}_c(z,\bz),\\
			&\he^{(1)} = u^{2}\frac{\D-2}{8}\D\e_{c}^{(-1)} +\frac{1}{2}u\,\D \e^{(0)}_c + \he^{(1)}_{c}(z,\bz),\\
			&\e^{(1)} = \e^{(1)}(u,z,\bz).
		\end{split}
	\end{equation}
	The same comments as before on the subleading integrations $ \e^{(n)},\he^{(n)} $ with $n\geq 1 $ can be repeated here. However, in this case we find two distinct arbitrary functions of the angular variables, $ \e_{c}^{(-1)} (z,\bz)$ and $\e^{(0)}_c(z,\bz) $, to which we can associate two independent contributions to the asymptotic charge, which correspond to the leading and the sub-leading soft photon theorem, as was derived in \cite{Campiglia:2016hvg}. 
	
	\subsubsection{\boldmath{$ \cO\left( r^N \right) $} asymptotic symmetries}
	As understood, we can in principle generalise the falloffs to an arbitrary $ r^N $ order ($ N>0 $) by taking $ a=\cO(r^N) $. Consequently, we consider $ \cO(r^N) $ gauge parameter, with expansion
	\begin{equation}\label{lorenz_par_N}
		\e = \sum_{n=-N}^{\infty} \frac{\e^{(n)}}{r^n} + \sum_{n=1}^{\infty} \he^{(n)} \frac{\log r}{r^n},
	\end{equation}
where the second series starts again from $ n=1 $.

The wave equation implies that, for every $ 0\leq n\leq N $
	\begin{equation}
		\e^{(-n)} = \sum_{m=0}^{N-n} u^{m} \e^{(-n,m)}
	\end{equation}
where the $ u $-independent part $ \e^{(-n,0)} $ is unconstrained at every order $ n $, since it comes from a $ u- $integration constant. Note that, although we used the same notation of \eqref{fur_exp}, this expansion is valid for every value of $ u $.

Consistently with the previous notation, we define 
\begin{equation}
	 \e_c^{(-n)}:=\e^{(-n,0)} .
\end{equation}
Given all the $ \e_c^{(-n)} $ for $  0\leq n\leq N  $, the wave equation completely determines the structure of the $ u-$dependent terms to be
\begin{equation}
	\e^{(-n,m)} =\frac{1}{2^{m}} \frac{n!}{ m! (n+m)! } \left[\prod_{i=1}^{m} \big(\D +(i+n)(i+n+1)\big)\right] \e^{(-n-m)}_c,
\end{equation}
where $ \e^{(-n-m)}_c=0 $ if $ n+m >N $. Hence, there are $ N $ independent free functions of the angular variables which give contributions to the asymptotic charges. 

\subsection{Charges}
\subsubsection{Leading and subleading electric charges}
The asymptotic electric charge is formally determined by
\begin{equation}
	Q_e = \lim_{t\to\infty}-\int_{\S_t} du dz d\bz\, \g_{z\bz} ( \pr_u - \pr_r) (r^{2} F_{ur} \e)
\end{equation}
and taking the finite part. The integral is ill-defined because there are both $ t-$ and $ u- $divergences. However, these divergences are understood as ambiguities of the presymplectic potential and are analysed in detail in \autoref{sec:ren}.

We start defining 
\begin{equation}\label{rho}
	\r := r^2 F_{ur}\, \e.
\end{equation}
Note that studying the $ \cO(t^{0}) $ term is not equivalent to studying the $ \cO(r^{0}) $ term in general, except when considering a $ \cO(r^{0}) $ asymptotic parameter. Indeed, for a $ \cO(r) $ parameter, we have
\begin{equation}
\r = r  \left [F_{ur}^{(2)} \e^{(-1)}\right ]+ \left [F_{ur}^{(3)} \e^{(-1)}+ F_{ur}^{(2)} \e^{(0)}\right ]+\dots 
\end{equation}
which implies
\begin{equation}
	(\pr_u-\pr_r )\r = r \left [ \e^{(-1)} \pr_u F_{ur}^{(2)}\right ]+\left[  \e^{(-1)}\pr_uF_{ur}^{(3)} +\pr_u(F_{ur}^{(2)} \e^{(0)})-F_{ur}^{(2)} \e^{(-1)}\right] +\dots
\end{equation}
while, substituting $ r=t-u $, we find 
\begin{equation}
	(\pr_u-\pr_r )\r = t \left[ \e^{(-1)} \pr_u F_{ur}^{(2)}\right]+\left[  \e^{(-1)}\pr_uF_{ur}^{(3)} +\pr_u(F_{ur}^{(2)} \e^{(0)})-\e^{(-1)}\pr_u \left (u\,F_{ur}^{(2)}\right )\right] +\dots\,.
\end{equation}

Using the expression for $ \e^{(0)} $ in \eqref{o(r)_as} as well as those for $ F_{ur} $  \eqref{fur}-\eqref{dfur}
and, substituting $ r=t-u $, we find up to some integration by parts at $\cO(t^{0})  $
\begin{equation}\label{qasl}
	Q_{e} =-\int du dz d\bz \g_{z\bz}\left(  \frac{1}{2} u \D \e^{(-1)}_c \pr_u D\cdot A^{(0)} + \e^{(0)}_c \pr_u D\cdot A^{(0)} \ \right) = Q_{e}^{(-1)}+Q_{e}^{(0)}
\end{equation}
which matches the soft part of \cite{Campiglia:2016hvg,Peraza:2023ivy}. In particular, the charge $ Q^{(-1)} $ is the one derived in the context of the subleading soft photon theorem in \cite{Lysov:2014csa}, thus indicating that the $ \cO(t^{0}) $ term is the correct one to consider. This approach differs, for instance, from the one proposed in \cite{Conde:2016csj} where the charge corresponding to the subleading soft photon theorem is not overleading as our $ Q^{(-1)} $ but the first subleading.

\subsubsection{$ \cO\left(r^N\right) $ charges}
Next, we determine the $ \cO(r^N) $ charge. For a $ \cO(r^N) $ parameter, $ \r $, defined in \eqref{rho}, has the following expansion
\begin{equation}
	\r = \sum_{n=-N}^{\infty}\frac{\r^{(n)}}{r^n}+\sum_{n=1}^{\infty}{\hat\r}^{(n)}\frac{\log r}{r^n}.
\end{equation}
We focus on the part containing positive powers of $ r $, namely the first series with $ -N\leq n\leq 0 $, which we rewrite as 
\begin{equation}
		\sum_{n=0}^{N}{r^n}\,{\r^{(-n)}}
\end{equation}
with 
\begin{equation}
	\r^{(-n)} = \sum_{k=0}^{N-n}F_{ur}^{(2+k)}\e^{(-n-k)}.
\end{equation}
To evaluate the charge we need to integrate
\begin{equation}
	 (\pr_u-\pr_r) (r^{n} \r ^{(-n)}) = r^{n} \pr_u \r ^{(-n)} -nr^{(n-1)}\r ^{(-n)} 
\end{equation}
so that the $ \cO(t^{0}) $ terms reads
\begin{equation}
	\sum_{n=0}^{N}\left ((-u)^{n} \pr_u \r ^{(-n)} -n(-u)^{(n-1)}\r ^{(-n)}\right )  =  \sum_{n=0}^{N}\big(\pr_u ((-u)^{n} \r ^{(-n)})\big)
\end{equation}

Hence, the charge at $ \cO(t^0) $ reads
\begin{equation}
	Q_{e} =  -\sum_{n=0}^{N} \int_{\S_t} du dz d\bz  \big(\pr_u ((-u)^{n} \r ^{(-n)})\big).
\end{equation}
As one can observe, there are divergences from the $ du $ integral, coming from powers in $ u $ of both $ \e $ and $ F_{ur} $. These divergences can be renormalized as well, leaving us with the $ \cO(u^{0}) $ term. We can now perform the limit for $ t\to\infty $ and find
\begin{equation}
		Q_{e, \, \text{ren}} = \sum_{n=0}^{N} Q_{e, \, \text{ren}}^{(-n)} =  \sum_{n=0}^{N} \int_{\cI^+_-} dz d\bz \,\g_{z\bz}\;  \e_c^{(-n)} F_{ur}^{(2+n,0)}\,,
\end{equation}
where the meaning of $  F_{ur}^{(2+n,0)} $ is explained in \eqref{fur_exp}. The renormalization discussion is presented in detail in \autoref{sec:ren} and the results match the ones discussed in \cite{Peraza:2023ivy}.

Finally, we can express everything as a function of the free data $ A^{(0)} $ by means of the residual gauge fixing and, using the expressions derived in \autoref{app:eom_sol}, we find 
\begin{equation}
\begin{split}
	\qquad Q_{e,\, \text{ren}}^{(0)} =& \int_{\cI^+_-} dz d\bz \,\g_{z\bz}\;  \e_c^{(0)} \D A^{(0,0)}\,.\\
	  Q_{e,\, \text{ren}}^{(n)} =&- \int_{\cI^+_-} dz d\bz \,\g_{z\bz}\;  \e_c^{(-n)} \D A^{(n,0)}\,. \qquad (n\geq 1)
\end{split}
\end{equation}

\subsubsection{Global charges in Lorenz gauge}\label{subs:glob}
Even though it is not strictly related to our purposes, let us add a few comments to the subject of the global charges, which should appear in
\begin{equation}\label{global}
	Q_{e}^{(0)} = \int dzd\bz \g_{z\bz} \e_c^{(0)} F_{ur}^{(2)}
\end{equation}
as the case of $ \e_c^{(0)} $ being a constant parameter. However, in Lorenz gauge we have $ F_{ur}^{(2)} = D\cdot A^{(0)}$ (also in the presence of sources\footnote{The condition $ F_{ur}^{(2)} = D\cdot A^{(0)} $ can be derived by the Lorenz gauge condition and therefore it is unaltered in the presence of sources. See the \autoref{app:eom} for the order-by-order equations.}), and therefore we find $ Q_{\text{const}}^{(0)} =0$, which would imply the absence of the standard electric charge.

This problem is solved by noting that the field strength falloffs, consistently with the equations of motion, allow for a
\begin{equation}
	A_r = \cO\left (\frac{1}{r}\right )
\end{equation}
with a constant $ A_r^{(1)} $. The correct expression for $ F_{ur}^{(2)} $, also in the presence of sources, would be 
\begin{equation}
	 F_{ur}^{(2)} = D\cdot A^{(0)}+A_r^{(1)},
\end{equation}
thus providing a term in \eqref{global} that survives in the constant $ \e^{(0)}_c $ case. 

Neglecting the constant $ A_r^{(1)} $ has no consequence in the analysis of asymptotic symmetries, characterized by non-trivial $ \e^{(n)}_{c} $, but it is crucial to find consistent global charges. To our knowledge, this detail has been first pointed out in \cite{Ferrero:2024eva}.

\section{Two-form}\label{sec:2f}
	In $ D=4 $, a two-form gauge field $ \cB_{\m\n} $ is dual, on-shell, to a scalar field $ \f $ through
	\begin{equation}\label{scalar_duality}
		H = *d\f,
	\end{equation}
	where we introduced the field strength $ H=d\cB $. By assuming the standard behaviour of a scalar field approaching null infinity  
	\begin{equation}
		\phi = \frac{\f^{(1)}(u,z,\bz)}{r}+\dots\,,
	\end{equation} 
	the duality induces the field strength falloffs, namely 
	\begin{equation}\label{constraints_H}
		H_{uri}=\cO(r^{-1}),\qquad  H_{uij} = \cO(r),\qquad  H_{rij} = \cO(1)\,.
	\end{equation}
	Accordingly to the setup chosen for the spin-one case, the field strength admits a power expansion, thus allowing for logarithms in the gauge field components only in pure gauge sectors. We work again in Lorenz gauge, namely 
	\begin{equation}
		\nabla^\m \cB_{\m\n} =0,
	\end{equation}
	so that the residual gauge parameter obeys to the Maxwell equations 
	\begin{equation}\label{B_Lorenz_gauge_1}
		\Box \l_\m-\nabla_\m \nabla\cdot \l = 0.
	\end{equation}
	Furthermore, $  \l_\m $ itself possesses a so-called gauge-for-gauge redundancy, namely 
	\begin{equation}
		\d_\e \l_\m = \pr_\m \e\,,
	\end{equation}
	with $ \e $ being a scalar function. We use $ \e $ to reach the Lorenz gauge for the parameter $ \l_\m $, so that it obeys to 
	\begin{equation}\label{Lor_par}
		\Box \l_\m= 0,\qquad \nabla\cdot \l = 0,
	\end{equation}
	and the residual gauge-for-gauge parameter obeys to 
	\begin{equation}
		\Box \e = 0.
	\end{equation}

\subsection{Falloffs and field strength}
	The two-form analogue of \eqref{A_expansion} is 
	\begin{equation}\label{B_expansion}
		\cB_{\m\n}= B_{\m\n}+\pr_{[\m}b_{\n]}=\sum_{n,n_1,n_2} \left [\frac{B_{\m\n}^{(n)}}{r^{n}} + \pr_{[\m}\left (b_{\n]}^{(n_1)}\frac{1}{r^{n_1}}+ {\hat b}^{(n_2)}_{\n]}\frac{\log r}{r^{n_2}}\right )\right ]\,,
	\end{equation}
	where the pure gauge part has been factored out, so that only the first series is relevant to the field strength.
	
	The field strength falloffs \eqref{constraints_H}, together with the ansatz about the power expansion, the equations of motion and the Lorenz gauge condition, determine the falloffs of the first series to be
	\begin{equation}\label{beta_fall}
		B_{ij} = \cO(r),\qquad B_{ui} = \cO(1),\qquad B_{ri} = \cO(r^{-1}),\qquad B_{ur} = \cO(r^{-2})\,.
	\end{equation}
	 In particular, $ B_{ij}^{(-1)} (u,z,\bz) $ has the role of the free data and we assume again that it approaches a well-defined function in the limit $ u\to-\infty $ and admits a $ u- $power expansion near $ \cI^+_- $ as in \eqref{free_scri}.
	
	 We can also write \eqref{B_expansion} as 
	\begin{equation}
		\cB_{\m\n} = \sum \frac{\cB_{\m\n}^{(n)}}{r^{n}} + \sum \hat{\cB}_{\m\n}^{(n)} \frac{\log r}{r^{n}}\,.
	\end{equation}
	As a consequence of the previous discussion and the equations of motion, we find that the field strength approaches $ \cI^{+}_{-} $ as
	\begin{equation}
		H_{uri} = \sum_{n=1}^{\infty}\frac{1}{r^n}\left[\sum_{m=0}^{n-1}u^m H_{uri}^{(n,m)} + 	h_{uri}^{(n)}(u,z,\bz)  \right],
	\end{equation}
	where $ h_{uri}^{(n)} \to 0 $ for $ u \to -\infty $. 
	
	In order to discuss asymptotic symmetries we apply the splitting previously used for $ A_i $ to the parameter component $ \l_i $, with components
	\begin{equation}\label{split_eps}
		\l_i = D_i \l+\ve_{ij}D^j \lp,
	\end{equation}
	and expansion coefficients $\l^{(n)},\hl^{(n)} $ and $ \l^{\prime(n)}, \hl^{\prime(n)}$ . By looking at \eqref{B_charge}, we observe that only the second component can contribute to the asymptotic charge, since
	\begin{equation}
		D^i H_{uri} = 0
	\end{equation} 
	as a consequence of the Bianchi identity $ \nabla_{[i}H_{urj]}=0 $. This can be understood by noting how the term $ D_i\l $ in $ \l_i $ resembles a gauge-for-gauge redundancy. Similar expansions can be proposed for the components 
	\begin{equation}\label{B_r-B_u}
	\begin{split}
			B_{ri} &= D_i B_r + \ve_{ij}D^j B_r^{\prime}\\
			B_{ui} &= D_i B_u + \ve_{ij}D^j B_u^{\prime},	
	\end{split}
	\end{equation}
	while, for the free data $ B_{ij} $ we write
	\begin{equation}\label{Bij}
		B_{ij} = \ve_{ij} B.
	\end{equation}
	Under a gauge transformation parametrised by $ \l,\lp $, we have
	\begin{equation}
		\d_{\l} B = 0,\qquad \d_{\lp} B = \D \lp,
	\end{equation}
	which further clarifies the different roles played by the two components of the splitting \eqref{split_eps}. 
	\subsection{Asymptotic symmetries}
	Given the expression \eqref{B_charge}, we observe that to find a non-vanishing charge, $ \l_i $ has to fall off at most as $ \cO(r) $, because of \eqref{constraints_H}. To find a non-trivial $ \cO(r) $ term in $ \lp $, it	must admit a polyhomogeneous expansion such as
	\begin{equation}\label{l_exp}
		\lp = \sum_{n=-{(N+1)}}^\infty\frac{\l^{\prime\, (n)}}{r^n}+\sum_{n=0}^\infty\hl^{\prime\, (n)}\frac{\log r}{r^n},
	\end{equation}
	with $ N\geq 0$. In particular, the second series is necessary to have a nontrivial $ \l^{\prime\,(-1)} $. 
	
	Furthermore, we observe how the splitting \eqref{split_eps} is particularly useful when working in Lorenz gauge. Indeed, the parameter, obeying to \eqref{Lor_par},
	completely decouples after the splitting. As detailed in \autoref{app:eom}, the equations $ \Box \l_u= 0,\; \Box \l_r = 0 $ and $ \nabla\cdot \l = 0 $ involve only the divergence of $ \l_i $ and therefore only $ \l $. At the same time, the equation for $ \Box \l_i $ decouples as well and, in particular, the order-by-order equations for $ \lp $ are 
	\begin{equation}\label{system_ep}
		\begin{cases}
				-2n \pr_u \l^{\prime\,(n)} + 2 \pr_u\hl^{\prime(n)}= [\D + n(n-1)]\l^{\prime\,(n-1)}-2(n-1)\hl^{\prime(n-1)}\, ,\\
					-2n \pr_u \hl^{\prime\,(n)} = [\D + n(n-1)]\hl^{\prime\,(n-1)}
		\end{cases}	
	\end{equation}
	where $ \hl^{(n)}=0 $ for $ n<0 $ in our expansion. This system is quite similar to the one for a scalar obeying the wave equation, up to a redefinition of the orders due to the fact that we are studying angular components $ \l_i $\footnote{\label{foot:par}Sometimes a different parametrization is chosen, so that a vector reads 
	$$ \l = \l^u \pr_u + \l^r \pr_r + \frac{1}{r}\l^i \pr_i\, , $$
	with the extra $ \tfrac{1}{r} $ in front of the angular component. With this parametrization, the system would have been identical to the scalar one.}. In particular, it is easy to check that, without a $ \hl^{\prime\,(0)} $, the system \eqref{system_ep} leads to 
\begin{equation}
	\D \l^{\prime\,(-1)} = 0\,,
\end{equation}
thus proving we cannot find asymptotic symmetries in Lorenz gauge without a polyhomogeneous expansion.

\subsubsection{\boldmath $ \cO(1) $ asymptotic symmetries}
	While it would be sufficient to consider only $ \lp $, as clarified  earlier, we start by briefly reviewing the result \cite{Ferrero:2024eva}, where the $ \cO(1) $ of a two-form asymptotic symmetries are understood as a component of the double-copy supertranslations. To this end, we consider
	\begin{equation}\label{2_lorenz_falloffs_1}
		\cB_{ur}= \cO(r^{-1}),\qquad \cB_{ui} = \cO(\log r),\qquad \cB_{ri} = \cO(1),\qquad \cB_{ij}=\cO(r),
	\end{equation}
	where the overleading terms respect to \eqref{beta_fall} are understood as pure gauge. Relative to the writing \eqref{B_expansion}, the falloffs \eqref{2_lorenz_falloffs_1} mean 
	\begin{equation}
		\begin{split}
				&B_{ij} = \cO(r),\qquad B_{ui} = \cO(1),\qquad B_{ri} = \cO(r^{-1}),\qquad B_{ur} = \cO(r^{-2}),\\
				&b_u = \cO(1),\qquad \; \; b_r = \cO(1),\qquad \; \;\, b_i = \cO(r)\,.
		\end{split}
	\end{equation}	
	Note in fact that it is not possible to have a non-trivial $ \cB_{ur}^{(1)} $ and $ \cB_{ri}^{(0)} $ without having $ \cB_{ui} = \cO(\log r) $, as observed in \cite{Ferrero:2024eva}.
	
	To find leading orders with a non-trivial angular dependence we require the gauge parameter to admit a polyhomogeneous expansion as well, in particular 
	\begin{equation}
		\begin{split}
			&\l_u = \sum_{n=0}^\infty\frac{\l_u^{(n)}}{r^n}+\sum_{n=1}^\infty\hl_u^{(n)}\frac{\log r}{r^n},\\ 
			&\l_r =  \sum_{n=0}^\infty\frac{\l_r^{(n)}}{r^n}+\sum_{n=1}^\infty\hl_r^{(n)}\frac{\log r}{r^n},\\
			&\l_i =  \sum_{n=-1}^\infty\frac{\l_i^{(n)}}{r^n}+\sum_{n=0}^\infty\hl_i^{(n)}\frac{\log r}{r^n}\,.
		\end{split}
	\end{equation}
	Equations \eqref{Lor_par} imply on the first terms that
	\begin{equation}
		\l_u^{(0)}=\l_u^{(0)}(z,\bz),\quad 	\l_r^{(0)}=\l_r^{(0)}(z,\bz),\quad 	\l_i^{(-1)}=\l_i^{(-1)}(z,\bz),\quad 	\hl_r^{(1)}=\hl_r^{(1)}(z,\bz)
	\end{equation}
	and fix a relation between the leading orders, namely 
	\begin{equation}
		\D \l_r^{(0)} +2(\l^{(0)}_u-\l^{(0)}_r)-2D^i\l_i^{(-1)}=0.
	\end{equation}
	We parametrise the solutions to this equation as ``supertranslations+something else'', namely\footnote{Let us remind that indices are raised as
	\begin{equation*}
		\l^{u\,(n)} = -\l^{(n)}_r,\quad	\l^{r\,(n)}=  \l^{(n)}_r-\l^{(n)}_u,\quad\l^{A\,(n)}=  \g^{ij}\l^{(n+2)}_j
	\end{equation*} 
	so that we recognise $ \cT(z,\bz) $ as the counterpart of the function that parametrises supertranslations in the gravitational case.}
	\begin{equation}
		\l_r^{(0)} = -\cT(z,\bz), \quad \l_i^{(-1)} = -D_i \cT + \cR_i(z,\bz), \quad \l^{(0)}_u = -\frac{\D+2}{2}\cT - D^i\cR_i.
	\end{equation}
	The supertranslations part here resembles a gauge-for-gauge redundancy. Indeed it can be fixed away by means of a gauge-for-gauge symmetry of a scalar parameter with $ \e^{(-1)} = \cT(z,\bz) $. The other part, parametrised by $ \cR_i $, can be decomposed again in 
	\begin{equation}
		\cR_i =  D_i \cR + \ve_{ij} D^j \cR^{\prime}
	\end{equation}
	and only $ \cR^{\prime} $ contributes to the charge, while the existence of $ \cR $ signals the possibility of performing a further residual gauge fixing.

\subsubsection{\boldmath $ O\left(r^{N}\right) $ asymptotic symmetries}
	Let us now discuss a $ \cO(r^N) $ asymptotic parameter, with the expansion \eqref{l_exp}. Using the splitting \eqref{split_eps}, the only contribution to the charge comes from the term $ \lp $. In fact, to study the asymptotic symmetries of a two-form we can consistently choose 
	\begin{equation}
		\l_u = 0,\qquad \l_r = 0,\qquad \l = 0.
	\end{equation}
	With this choice we cannot appreciate the possible double-copy structure as in the previous discussion, but it is the simplest way to analyse the two-form asymptotic symmetries. This setup implies, for instance, that the falloffs of $ B_{ur}, B_{r},B_{u} $, where we used the notation introduced in \eqref{B_r-B_u}, can be always preserved. The falloffs we need to generalise by means of pure gauge terms are only $ B, B_r^{\prime}, B_u^{\prime} $. In particular, $ \lp $ acts on these components as
	\begin{equation}
	\begin{split}
			&[ \d B ]^{(n)} = \D\l^{\p\,(n)},\quad 	[ \d B^{\prime}_u ]^{(n)} =\pr_u\l^{\p\,(n)},\quad 	[ \d B^{\prime}_r ]^{(n)} = -(n-1)\l^{\p\,(n-1)}+\hl^{\p\,(n-1)}\\
			&\widehat{[ \d B ]}^{(n)} = \D\hl^{\p\,(n)},\quad 	\widehat{[ \d B^{\prime}_u ]}^{(n)}  =\pr_u\hl^{\p\,(n)},\quad 	\widehat{[ \d B^{\prime}_r ]}^{(n)} = -(n-1)\hl^{\p\,(n-1)}
	\end{split}
	\end{equation}
	
The wave equation for $ \lp $, which is reported order-by-order in \eqref{system_ep}, completely fixes the $ u- $dependence on $ \cI^+ $ being again of the form 
	\begin{equation}
		\l^{\prime\,(-n)} = \sum_{m=0}^{N-n} u^{m} \l^{\prime\,(-n,m)}\,,
	\end{equation}
where we are now focusing on the $ n\geq1$ part, which gives non-vanishing contributions to the charge, and in particular 
	\begin{equation}
			\l^{\prime\,(-n,m)} =\frac{1}{2^{m}} \frac{(n-1)!}{ m! (n+m-1)! } \left[\prod_{i=1}^{m} \big(\D +(i+n-1)(i+n)\big)\right] \l^{\prime\,(-n-m)}_c\,,
	\end{equation}
where the result is identical the case of the scalar wave equation except for a shift $ n\to n-1 $ due to the fact that the equations of motion of the angular component change the $ r $-order (see \autoref{foot:par}).

\subsection{Charges}	
	\subsubsection{Leading and subleading charges.}
	 We study the limit $ t\to \infty $ of 
	\begin{equation}
	-	\int_{\S_t} du dz d\bz \g_{z\bz} ( \pr_u - \pr_r) (\g^{ij} H_{uri} \l_{j})\,.
	\end{equation}
	and consider the $ \cO(t^{0}) $ term. We rewrite the integral using  the splitting \eqref{split_eps} and integrating by parts as
	\begin{equation}
		\int_{\S_t} du dz d\bz\g_{z\bz} ( \pr_u - \pr_r) \r_{H}\,,
	\end{equation}
where we define
\begin{equation}
	\r_{H} := \lp \ve^{ij}  D_j H_{uri}\, .
\end{equation}  

We start considering the leading and subleading contributions to the asymptotic charge, with a $ \lp=\cO(r^{2}) $. In particular, we have
	\begin{equation}
		\begin{split}
			(\pr_u -\pr_r)\r_{H} =& r\left [\l^{\p\,(-2)} \ve^{ij} \pr_u D_jH_{uri}^{(1)}\right ]+r^{0}\left [-\l^{\p\,(-2)}\ve^{ij}D_jH_{uri}^{(1)} \right.\\ &\left.+\,\pr_u (\l^{\p\,(-1)}\ve^{ij} D_{j}H_{uri}^{(1)} )+\l^{\p\,(-2)}\ve^{ij}\pr_u D_jH_{uri}^{(2)} \right]+\dots
		\end{split}
	\end{equation}
and the $ \cO(t^{0}) $ term is given by
\begin{equation}
	\pr_u \left(- u\l^{\p\,(-2)}\ve^{ij} D_jH_{uri}^{(1)}+\l^{\p\,(-1)}\ve^{ij}D_jH_{uri}^{(1)}+\l^{\p\,(-2)}\ve^{ij}D_jH_{uri}^{(2)} \right),
\end{equation}
while the term linear in $ t $ is neglected by means of symplectic renormalization.

To evaluate this, we need the expressions
	\begin{equation}
	\begin{split}
			&H_{uri}^{(1)} = -D^jB_{ij}^{(-1)}\\
			&\pr_uH_{uri}^{(2)} = \frac{\D-1}{2}D^jB_{ij}^{(-1)},
	\end{split}
	\end{equation}
and, by means of \eqref{Bij}, we find
\begin{equation}
	\begin{split}
		&\ve^{ij}D_j H_{uri}^{(1)} = \D B^{(-1)},\\
		& \ve^{ij} \pr_u D_j H_{uri}^{(1)} = -\frac{\D^{2}}{2} B^{(-1)}.
	\end{split}
\end{equation}
Finally, using
\begin{equation}\begin{split}
	&\l^{(-2)} = \l^{(-2)}_{c} (z,\bz),\\
	&\l^{(-1)} =  u\frac{\D+2}{2} \l^{\p\,(-2)}_{c}(z,\bz)+\l^{\p\,(-1)}_{c}(z,\bz),
	\end{split}
\end{equation}
we arrive to
\begin{equation}
	Q_{B} = Q_{B}^{(0)}+Q_{B}^{(-1)}\,,
\end{equation}
with 
\begin{subequations}
	\begin{align}
		&Q_{B}^{(0)} = \int du dz d\bz \g_{z\bz}\, \l_{c}^{\p \,(-1)} \pr_u \D B^{(-1)} ,\\
		\label{qbsl}& Q_{B}^{(-1)} = \int du dz d\bz \g_{z\bz} \frac{u}{2} \l_{c}^{\p \,(-2)}  \pr_u \D^{2} B^{(-1)}.
	\end{align}
\end{subequations}

The charge $ Q^{(0)}_{B} $ matches the one presented in \cite{Ferrero:2024eva}. It corresponds to the leading soft scalar theorem, originally derived in \cite{Campiglia:2017dpg}, as studied in \cite{Campiglia:2018see,Francia:2018jtb}. The charge $ Q^{(-1)}_{B} $, to which we refer as subleading, is presented here for the first time. Interestingly, it has a form identical to its electromagnetic counterpart $ Q^{(-1)}_e $ derived in \eqref{qasl} which, up to an integration by parts, reads
	\begin{equation}
		Q^{(-1)}_{e}=		\int du dz d\bz \g_{z\bz}  \frac{u}{2}   \e^{(-1)}_c \pr_u \D^2 A^{(0)}
	\end{equation}
	where we used the splitting \eqref{split_A} to highlight the similarity between the two charges. Our guess is that the conservation of the charge \eqref{qbsl}, up to the identification of $ B^{(-1)} $ with the scalar mode $ \f^{(1)} $, might correspond to a subleading version of the soft scalar theorem presented in \cite{Campiglia:2017dpg}. 
	
	To our knowledge, however, no subleading formulation of the soft scalar theorem of \cite{Campiglia:2017dpg} has been derived yet. Nonetheless, we expect the subleading soft scalar factor to have a similar structure to the photon and graviton analogues. This is motivated, for instance, by the structure of subleading soft factors for dilatons, derived in \cite{DiVecchia:2015jaq,DiVecchia:2015oba}. 
	
	Furthermore, we notice the proposal in (3.43) of \cite{Campiglia:2017dpg} for the smeared version of the scalar charges is also motivated by the analogy with electromagnetism and gravity. In this sense, the natural proposal for a subleading version of the charge would be our $ Q^{(-1)}_B $.

\subsubsection{\boldmath $ O\left (r^{N}\right ) $ charges.}
	Let us now consider the $ \cO(r^N) $ charges with arbitrary $ N $ by repeating the same step of the previous case. We have
	\begin{equation}
		\r_{H} = \sum_{n=-N}^{\infty} \frac{\r_{H}^{(n)}}{r^n}+\sum_{n=1}^{\infty} {\hat \r}^{(n)}\frac{\log r}{r^n}\,. 
	\end{equation}
We focus on the part with $ -N\leq n\leq 0 $ with 
\begin{equation}
	\r_H^{(-n)} = \sum_{k=0}^{N-n}\l^{\p\,(-n-k-1)}\ve^{ij} D_jH_{uri}^{(1+k)}\,,
\end{equation}
	so that the $ \cO(t^{0}) $ term reads
	\begin{equation}
		 \sum_{n=0}^{N}\pr_u \left ((-u)^{n} \r ^{(-n)}\right ), 
	\end{equation}
with the charge 
\begin{equation}
	Q_{B} = \sum_{n=0}^{N}\int_{\S_t} du dz d\bz \g_{z\bz} \, \pr_u \left((-u)^{n} \r ^{(-n)}\right).
\end{equation}
There are again $ u- $divergences, but they are understood in terms of ambiguities of the presymplectic potential. We are left only with the $ u- $independent terms and a well-defined integral, so that we can perform the limit $ t\to\infty $ and obtain
\begin{equation}
	Q_{B,\,\text{ren}} = \sum_{n=0}^{N} Q^{(-n)}_{B,\,\text{ren}} = -\int_{\cI^+_-} dz d\bz \g_{z\bz} \,\l^{(-(n+1))}_{c} \ve^{ij}D_j H_{uri}^{(n+1,0)}.
\end{equation}

The confrontation with the spin-one counterpart is straightforward: we find $ N $ independent asymptotic charges, parametrised by $ N $ arbitrary scalar functions of the angular variables, arising from $ u-$integrations. Using the expressions obtained in \autoref{app:eom_sol} we find 
	\begin{equation}\label{charges_B}
		\begin{split}
			\qquad Q_{B}^{(\,0\,)} \;=& -\int_{\cI^+_-} dz d\bz \g_{z\bz} \l^{\prime(-1)}_{c} \D B^{(-1,0)}\,,\\
				\qquad Q_{B}^{(-1)} =& -\int_{\cI^+_-} dz d\bz \g_{z\bz} \l^{\prime(-2)}_{c} \frac{\D^2}{2} B_r^{\p(1,0)}\,,\\
		\qquad	Q_{B}^{(-n)} =&\frac{n-1}{2} \int_{\cI^+_-} dz d\bz \g_{z\bz} \l^{\prime(-n-1)}_{c} \D B^{(n-1,0)}\,\, \qquad (n\geq 2)\,.
		\end{split}
\end{equation}

\section{Renormalization of the charge}\label{sec:ren}
Our study of asymptotic charges encounters two types of divergences: those arising from positive powers of 
$ t $ and those resulting from the $ du $ integration over positive powers of $ u $. To address these issues, we employ a symplectic renormalization procedure, as proposed in \cite{Freidel:2019ohg}.

The presymplectic potential admits two types of ambiguities that allow us to define an equivalence relation
\begin{equation}\label{ambig_2}
	\th^\m \sim \th^\m + \pr_\n\Upsilon^{\m\n} +\d\Xi ^\m,
\end{equation}
with $ \Upsilon^{\m\n} $ being antisymmetric. Our prescription is to use these ambiguities to cancel the divergences while leaving the rest unchanged.

We do not need to specify the theory to which we are referring, since all the expansions and assumptions apply to both one- and two-form cases, as we prove. This is a consequence of using the same setup for both theories. In $ (u,t,z,\bz) $ coordinates, we have 
\begin{equation}\label{dth_t}
	\pr_t \th^{t} \approx \d \cL -\pr_u\th^{u} - \pr_i \th^{i},
\end{equation}
where ``$ \approx $'' denotes an on-shell relation. As $ t\to+\infty $, $ \th^{t} $ has the general structure
\begin{equation}\label{exp_pres}
	\th^{t} = \sum_{n=1}^{N} t^{n} \th^{t}_{(n)} (u,z,\bz) + \th^{t}_{(0)}(u,t,z,\bz),
\end{equation}
where $\th^{t}_{(0)}(u,t,z,\bz) $ is such that $ \lim\limits_{t\to\infty} \th^{t}_{(0)}(u,t,z,\bz) = \th^{t}_{(0)}(u,z,\bz)$. In particular, logarithmic terms such as $t^{-n} \log t $ are inside $ \th^{t}_{(0)}(u,z,\bz) $ and, therefore, not relevant. Similar expansions are introduced for $ \cL,\th^{u},\th^{i} $ with respective coefficients $ \cL_{(n)},\th^{u}_{(n)},\th^{i}_{(n)} $ . 

Let us verify that the expansion \eqref{exp_pres}  holds in both one- and two-form cases. For the spin-one case, we write
\begin{equation}
	\th^{t}_{e} = \th^{u}_{e}+\th^{r}_{e} = r^2 \g_{z\bz}(F^{u\n}+F^{r\n})\d\cA_{\n}\,.
\end{equation}
In our setup, the field strength admits a power expansion, while for the gauge field variation we have
\begin{equation}
	\d \cA_\n = \d A_\n - \pr_\n \d a ,
\end{equation} 
where $ A_\n $ admits a power expansion as well, while for $ a $ we chose to consider a polyhomogeneous expansion. In particular, we used 
\begin{equation}
	a = \sum_{n=-N}^{\infty} \frac{a^{(n)}}{r^n} + \sum_{n=1}^{\infty} \ha^{(n)} \frac{\log r}{r^n},
\end{equation} 
with the logarithmic series starting at order $ \cO\left (\frac{\log r}{r} \right) $. Summing over indices we find 
\begin{equation}\label{th_t_exp}
	\th^{t}_{e} = r^2 \g_{z\bz}\left[F_{ur}(\d\cA_{u} - \d\cA_{r})-\frac{\g^{ij}}{r^2}F_{uj}\d \cA_i \right]
\end{equation}
and, using the field strength falloffs \eqref{phys}, we deduce that the first logarithmic term in $ \th^{t}_{e} $ is of order $ \cO\left(\frac{\log r}{r}\right) $ and therefore of order $ \cO\left(\frac{\log t}{t}\right) $. The highest order, instead, is of order $ \cO(r^N) $ and therefore $ \cO(t^N) $.
 
Let us repeat the analysis for the two-form case. We have 
\begin{equation}\label{th_B^t}
		\begin{split}
			\th^{t}_{B} &= r^2 \g_{z\bz} (H^{u\m\n}+H^{r\m\n})\d \cB_{\m\n} \\
			&= \g_{z\bz} \left[\g^{ij}H_{uri}(\d \cB_{uj}-\d\cB_{rj})-\frac{\g^{ik}\g^{jl}}{r^2}H_{uij}\d \cB_{kl}\right].
		\end{split}
\end{equation} 
Field strength components admit a power expansion with falloffs \eqref{constraints_H}. The two-form gauge field variation is 
\begin{equation}
	\d\cB_{\m\n}=\d B_{\m\n}-\pr_{[\m}\d b_{\n]}
\end{equation}
with $ b $ admitting a polyhomogenoues expansion of the form 
\begin{equation}
	\begin{split}
			&b_u = \sum_{n=-N}^\infty\frac{b_u^{(n)}}{r^n}+\sum_{n=1}^\infty{\hat{b}}_u^{(n)}\frac{\log r}{r^n},\\ 
			&b_r =  \sum_{n=-N}^\infty\frac{b_r^{(n)}}{r^n}+\sum_{n=1}^\infty{\hat{b}}_r^{(n)}\frac{\log r}{r^n},\\
			&b_i =  \sum_{n=-N-1}^\infty\frac{b_i^{(n)}}{r^n}+\sum_{n=0}^\infty{\hat{b}}_i^{(n)}\frac{\log r}{r^n}\,.
	\end{split}
\end{equation}
Substituting these in the equation\eqref{th_B^t}, we find exactly the expansion \eqref{exp_pres}. 

The equation \eqref{dth_t} implies that every $  \th^{t}_{(n)}  $ with $ n>0 $ corresponds to an ambiguity, namely
\begin{equation}
	\th^{t}_{(n)} = \frac{1}{n}\left (\d\cL_{(n-1)}-\pr_u\th^{u}_{(n-1)} - \pr_i \th^{i}_{(n-1)}\right )\,.
\end{equation}
Therefore, we can identify
\begin{equation}
\begin{split}
		\Upsilon^{tu} =& \sum_{n=1}^{N} \frac{1}{n}t^{n} \th^{u}_{(n-1)} +\Upsilon^{tu}_{(0)}\,,\\
		\Upsilon^{ti} =& \sum_{n=1}^{N} \frac{1}{n}t^{n} \th^i_{(n-1)} +\Upsilon^{ti}_{(0)}\,,\\
		\Xi^{t} =& \sum_{n=1}^{N} -\frac{1}{n}t^{n} \cL_{(n-1)} +\Xi^{t}_{(0)}\,,
\end{split}
\end{equation}
where $ \Upsilon^{tu}_{(0)},\Upsilon^{tA}_{(0)},\Xi^{t}_{(0)} $ are (free) functions with a well-defined limit for $ t\to\infty $. We thus find that 
\begin{equation}
	\th^{t}\sim \th^{t}_{(0)}+ \pr_u\Upsilon^{tu}_{(0)}+\pr_i \Upsilon^{ti}_{(0)}+\d\Xi^{t}_{(0)},
\end{equation}
implying that only the $ t^{0} $ order contributes to the charge in the limit $ t\to\infty $. 

Next, we address the $ u- $divergences, which arise from the $ du $ integration. This is managed using $ \Upsilon^{t\m}_{(0)} $. Near $ u\to-\infty $, $ \th^{t}_{(0)} $ reads
\begin{equation}\label{th-uexp}
	\th^{t}_{(0)} = \sum_{n=1}^{N}\th^{t}_{(0,n)}u^{n-2}+\Th^{t}_{(0)}(u,z,\bz),
\end{equation}
where we factored out the part that is not $ u- $integrable, absent for $ N=0 $, while $ \Th^{t}_{(0)}(u,z,\bz) $ admits a $ u- $power expansion with leading order $ \cO(u^{-2}) $. 

To prove this expansion, we start from the setup used for $ \cO(1) $ asymptotic symmetries. The $ \cO(r^N) $ case would differ from the $ \cO(1) $ only for overleading powers of $ u $. Let us begin with electromagnetism and determine the $ \cO(t^{0}) $ term. The only contribution is from the last term in \eqref{th_t_exp} and reads
\begin{equation}
	\th^{t}_{e(0)} = \g_{z\bz} \g^{ij} \pr_u A_i^{(0)}\d \cA_j^{(0)},
\end{equation}
where we substituted $ F_{ui}^{(0)}= \pr_u A_i^{(0)} $. Near $ \cI^+_- $ we use the ansatz \eqref{free_scri} for $ A_i^{(0)} $ which, for $ u \to-\infty$, means
 	\begin{equation}
 		\pr_u A_i^{(0)} = \cO(u^{-2})\,.
 	\end{equation}
 Consistently with this assumption, we find that  $ \d \cA_j^{(0)} $ tends to a well-defined function of the angles for $ u\to-\infty $. Combining these two results, we find a $ u- $integrable quantity for $ \th^t_{(0)} $ in the $ N=0 $ case. Considering higher order terms with arbitrary $ N $, the setup differs only for $ a=\cO(r^N) $, therefore we are adding overleading powers of $ u $, finding non-integrable quantities as in the first series of the expansion \eqref{th-uexp}. 
 
Let us repeat the same computation for the two-form case in the $ N=0 $ setup. We have
\begin{equation}
	\th^{t}_{B(0)} = \g_{z\bz} \g^{ik}\g^{jl} \pr_u B_{ij}^{(-1)}\d \cB_{kl}^{(-1)},
\end{equation}
and near $ \cI^+_- $
\begin{equation}
	\begin{split}
		 &B_{ij}^{(-1)} = B_{ij}^{(-1,0)}(z,\bz)+\cO\left (u^{-1}\right )\\
		 & \pr_u B_{ij}^{(-1)} = \cO\left (u^{-2}\right ).
	\end{split}
\end{equation}
Hence, the presymplectic potential in the $ N=0 $ case is integrable and order $ \cO(u^{-2}) $. Modifying the setup to arbitrary $ N $ adds positive powers of $ u $.

We now use $  \Upsilon^{t\m}_{(0)} $ to cancel the first series. This can be achieved, for instance, employing these expansions
\begin{equation}
	 \Upsilon^{t u }_{(0)} = \sum_{n=3}^{N+1} u^{n-2} \Upsilon^{t u }_{(0,n)},\qquad \Upsilon^{t i }_{(0)} = \frac{1}{u} \Upsilon^{t i }_{(0,-1)}
\end{equation}
with 
\begin{equation}
	\Upsilon^{t u }_{(0,n)} = -\frac{\th^{t}_{(0,n+1)}}{n-2},\qquad \pr_i  \Upsilon^{t i }_{(0,-1)} =-\th^{t}_{(0,1)}.
\end{equation}
This solution is not unique and depends on the specific structure of the renormalizing terms. Here we are assuming that $ \Upsilon^{t u }_{(0)} $ has a $ u- $power expansion, making it impossible to cancel a $ \tfrac{1}{u} $ term with a $ u- $derivative, which is why we used $ \pr_i  \Upsilon^{t i }_{(0,-1)} $.

Let us comment on the fact that the expansion \eqref{th-uexp} is related to the ansatz \eqref{free_scri}, specifically to the assumption that the free data approaches $ \cI^+_- $ with a $ u- $power expansion. As mentioned in \autoref{foot:tree}, other expansions can be considered. This is the case of \cite{Peraza:2023ivy}, for which the free data approaches a well defined function $ A_i^{(0,0)} $ faster that any $ u^{-n} $. This choice, however, would not alter equation \eqref{th-uexp},  but only the falloff of the integrable quantity, preserving the validity of the renormalization process. A different expansion is proposed in \cite{Alessio:2024onn}, namely
\begin{equation}
	A_i^{(0)} = 	A_i^{(0,0)}+ \sum_{n=1}^{\infty}\hA_{i}^{(0,n)} \frac{(\log u)^{n-1}}{u^{n}} +\dots \,,
\end{equation}
which instead would modify the equation \eqref{th-uexp}, and therefore would lead to some differences in the renormalization procedure, whose analysis we leave to a future work.

Ultimately, this process leaves us with well-defined quantities, thus showing that it is possible to renormalize the presymplectic potential to make the asymptotic charge finite while accommodating arbitrary powers of the asymptotic parameter. 

The symplectic renormalization procedure is a powerful tool that allow us to study higher order asymptotic symmetries and deal with divergences. This approach suggests the possibility of discovering an infinite set of asymptotic charges, broadly meaning an infinity of infinitely-many conserved charges. It is crucial, however, to emphasize that we used a specific prescription: to cancel the divergences by means of ambiguities while leaving unaltered the finite parts, which enter in the definition of asymptotic charges. However, it is clear that there are residual ambiguities that might further constraint the number of physical asymptotic charges. We could find more general forms for $ \Upsilon $ and $ \Xi $ to cancel some finite parts of $ \th^{t} $. In particular, we did not use the residual term $ \d \Xi_{(0)}^{t} $. 
		
\section{Conclusions}
In this work, we analysed $ \cO(r^N) $ one- and two-form asymptotic symmetries. We began by revisiting electromagnetism, which was studied in detail in \cite{Peraza:2023ivy}. We focused only on the soft part of the charges, with the aim of confronting it with the two-form ones, and we chose a slightly different setup w.r.t.\ \cite{Peraza:2023ivy}, mostly concerning to the expansion of the quantities involved. At any rate, our results are in agreement with those of \cite{Peraza:2023ivy}, in all places where comparisons were possible. We then moved to $ \cO(r^N) $ two-form asymptotic symmetries. Up to the symplectic renormalization procedure, we found $ N $ independent asymptotic charges, with each charge parametrised by an arbitrary function of the angles. The result for the $ \cO(r) $ case is particularly interesting, since it is the charge one would naively expect for the subleading soft scalar theorem in the spirit of \cite{Campiglia:2017dpg}. 

These results leave space for several future investigations. First, it would be interesting to repeat our analysis considering sources of various form, thus providing the hard part of the asymptotic charges. As mentioned, we think that this would require particular attention in the falloffs analysis. Nonetheless, let us remark that near spatial infinity, the interacting theory should be well-described by the free theory. In this limit, we expect the charge expressions \eqref{charges_B} to be valid in the interacting theory as well.

Another enticing perspective is to investigate higher-order asymptotic symmetries in other gauges. In this respect, let us remark that $ \cO(1) $ two-form asymptotic symmetries were analysed in radial gauge in \cite{Francia:2018jtb}, providing the surprising result of a vanishing charge. Our interpretation of this fact is that, in order to reach the radial gauge condition $ B_{r\m}=0 $, we employ a physical part of the gauge parameter $\l_{\m} $, which would contribute to the asymptotic charge. In fact, let us observe that we are able to find instead a nonvanishing asymptotic charge if working in ``subleading radial gauge'', namely
	\begin{equation}
		B_{r\m} = B_{r\m} (u,z,\bz)
\end{equation}
exactly. In this case, in fact, we gauged away only the subleading part of $ B_{r\m} $, leaving a $ \cO(r) $ residual gauge parameter, which contributes to the charge $ Q^{(0)}_{B} $. 
To our knowledge, higher-order asymptotic symmetries in electromagnetism were discussed only in Lorenz gauge. Let us propose here a possible extension to the radial gauge condition, which is reached imposing
\begin{equation}
	A_r =0\,,
\end{equation} 
leaving a $ \cO(1) $ gauge parameter $ \e(u,z,\bz) $. Then, the $ u- $dependent part of $ \e $ is typically employed to reach the retarded radial gauge, characterised by the extra condition $ A_u|_{\cI^+} =0$. Our opinion is that, to impose retarded radial gauge we are using a physical part of $ \e $, corresponding to all the subleading charges. If we consider only the radial gauge condition, we can for instance assume a $ u- $expansion for $ \e $
\begin{equation}
	\e^{(0)} = \sum_{n=0}^N u^n \e^{(0,n)} + \text{ subleading in }u\,.
\end{equation}
This parameter would lead to $ u- $divergences in the charge, which require to be renormalized, but also to some new finite terms that could correspond to the subleading asymptotic charges $ Q_e^{(n)} $. For gravity, $ \cO(r^2) $ asymptotic symmetries were studied also in Bondi gauge in \cite{Horn:2022acq}. 
In this respect, let us observe that the Lorenz gauge is natural from a double-copy viewpoint \cite{Anastasiou:2014qba} and therefore, given the result \cite{Ferrero:2024eva}, the comparison between the one- and two-form results is quite natural in this gauge.

Another possible generalisation involves exploring arbitrary spacetime dimension $ D $ and generic $ p-$forms, with the aim of exploring the duality between $ p- $ and $ (D-p-2)-$forms from the perspective of asymptotic symmetries. 

Additionally, we seek to better understand the potential connection between the $ \cO(r^n)$ asymptotic charge and the sub$^n $-leading soft theorem. In this respect, the existing duality between the two-form and a scalar in $ D=4 $ is particularly appealing, given the relative simplicity of scalar theories. To discuss this connection, we need to deal with the interacting theory, and therefore the inclusion of sources becomes crucial. Furthermore, this would also help in clarifying the role of the residual ambiguities. A deeper understanding of these terms is in fact crucial, as it would help to determine the physical significance of asymptotic symmetries at every order and to explore the limitations of symplectic renormalization.

\acknowledgments 
I would like to thank Andrea Campoleoni, Dario Francia, Carlo Heissenberg and Federico Manzoni for discussions and valuable comments. 

\appendix
	\section{Equations of motion}\label{app:eom}
	Let us remind that, for radial expansions, we employ the following notation
	\begin{equation}
		\vf = \sum_{n_1} \frac{\vf^{(n_1)}}{r^{n_1}}+\sum_{n_2} {\hat \vf}^{(n_2)} \frac{\log r}{r^{n_2}},
	\end{equation}
	so that $ \vf^{(n_1)} $ is the coefficient for $ r^{-n_1} $.

	\subsection{Order by order tricks }
	In evaluating order-by-order equations, these are the relevant terms that often occur:
	\begin{equation}\label{tricks}
		\begin{split}
			&\frac{1}{r^{\a}}\pr_r \vf : \quad \begin{cases}
				\big[\tfrac{1}{r^{\a}}\pr_r \vf\big]^{(n)} = {\hat \vf}^{(n-(\a+1))}-(n-(\a+1))\vf^{(n-(\a+1))}\\
				\big[\widehat{\tfrac{1}{r^{\a}}\pr_r \vf}\big]^{(n)} = -(n-(\a+1)){\hat \vf}^{(n-(\a+1))}
			\end{cases}\\
			&\;\;\,\pr^2_r \vf\;: \quad \begin{cases}
				\big[\pr^2_r \vf\big]^{(n)} = 	(n-1)(n-2)\vf^{(n-2)}-(2n-3){\hat \vf}^{(n-2)}\\
				\big[\widehat{\pr^2_r \vf}\big]^{(n)} = (n-1)(n-2){\hat \vf}^{(n-2)}
			\end{cases}
		\end{split}
	\end{equation}

	\subsection{Scalar }
	The scalar wave equation reads
	\begin{equation}
		\Box \e  = \left(\pr_r^2-2\pr_u\pr_r -\frac{2}{r}(\pr_u-\pr_r) + \frac{\D}{r^2}\right)\e=0
	\end{equation}
	which, order-by-order, implies the system
	\begin{equation}
		\begin{cases}
			(2-2 n) \partial_u \epsilon^{(n)}+2 \partial_u \he^{(n)}  =[\Delta+(n-1)(n-2)] \epsilon^{(n-1)}+(3-2 n) \he^{(n-1)}\,, \\
			(2-2 n) \partial_u \he^{(n)}  =[\Delta+(n-1)(n-2)] \he^{(n-1)}\,.
		\end{cases}
	\end{equation}

	\subsection{Vector}
A vector in Lorenz gauge obeys to
\begin{subequations}
	\begin{align}
		&\Box \cA_u = \left(\pr_r^2-2\pr_u\pr_r-\frac{2}{r}(\pr_u-\pr_r)+\frac{\D}{r^2}\right)\cA_u=0\,,\\
		&\Box \cA_r = \left(\pr_r^2-2\pr_u\pr_r-\frac{2}{r}(\pr_u-\pr_r)+\frac{\D}{r^2}\right)\cA_r+\frac{2}{r^2}(\cA_u-\cA_r)-\frac{2}{r^3}D\cdot A=0 \,,\\
		&\Box \cA_i = \left(\pr_r^2-2\pr_u\pr_r+\frac{\D-1}{r^2}\right)\cA_i-\frac{2}{r}D_i(\cA_u-\cA_r)=0\,,\\
		&\nabla\cdot \cA = -\pr_u\cA_r+\left(\pr_r+\frac{2}{r}\right)(\cA_r-\cA_u)+\frac{1}{r^2}D\cdot \cA=0\,.
	\end{align}
\end{subequations}
The gauge condition reads
\begin{equation}
	\begin{cases}
		\pr_u \cA_r^{(n)}-(n-3)\left(\cA_u^{(n-1)}-\cA_r^{(n-1)}\right)-D \cdot \cA^{(n-2)}+{\hat \cA}_u^{(n-1)}-{\hat \cA}_r^{(n-1)}=0\,,\\
		\pr_u {\hat \cA}_r^{(n)}-(n-3)\left({\hat \cA}_u^{(n-1)}-{\hat \cA}_r^{(n-1)}\right)-D \cdot {\hat \cA}^{(n-2)}=0\,,
	\end{cases}
\end{equation}
while the equations of motions are\footnote{Moving to the right the l.h.s.\ of each equation we recover the order-by-order expression of $ \left[ \Box \cA_\m \right]^{(n+1)} $.}
\begin{equation}
	\begin{aligned}
		&\begin{cases}
			(2-2n)\pr_u \cA_u^{(n)}+2\pr_u{\hat \cA}_u^{(n)}=[\Delta + 	(n-1)(n-2)]\cA_u^{(n-1)}-(2n-3){\hat \cA}_u^{(n-1)}\,,\\
			(2-2n)\pr_u {\hat \cA}_u^{(n)}=[\Delta + 	(n-1)(n-2)]{\hat \cA}_u^{(n-1)}\,,
		\end{cases}\\
		&\begin{cases}		
				(2-2n)\pr_u \cA^{(n)}_r +2\pr_u{\hat \cA}_r^{(n)}= 	[\Delta+n(n-3)]\cA_r^{(n-1)} + 2 \cA_u^{(n-1)}-2 D\cdot \cA^{(n-2)} -{ (2n-3)}{\hat \cA}_r^{(n-1)},\\	
			(2-2n)\pr_u {\hat \cA}^{(n)}_r = 	[\Delta+n(n-3)]{\hat \cA}_r^{(n-1)} + 2\, {\hat \cA}_u^{(n-1)}-2 D\cdot {\hat \cA}^{(n-2)}\,,		
		\end{cases}\\
		&\begin{cases}
			-2n\,\pr_u \cA_i^{(n)}+2\pr_u {\hat \cA}_i^{(n)}=[\Delta+n(n-1)-1]\cA_i^{(n-1)}-2D_i(\cA_u^{(n)}-\cA_r^{(n)})-(2n-1){\hat \cA}_i^{(n-1)}\,,\\
			-2n\,\pr_u {\hat \cA}_i^{(n)}=[\Delta+n(n-1)-1]{\hat \cA}_i^{(n-1)}-2D_i({\hat \cA}_u^{(n)}-{\hat \cA}_r^{(n)})\,.
		\end{cases}
	\end{aligned}
\end{equation}
These equations, containing the logarithmic part, are important for the two-form parameter. However, in order to establish the explicit relation between the field strength and the spin-one field, we need the equations only for the $ A-$part of $ \cA $, which admits a power expansion. These equations are identical to the equations of the pure logarithmic part, but here we report them for simplicity
\begin{subequations}
	\begin{align}
		\label{eq:gaugeA}&\pr_u A_r^{(n)}=(n-3)\big(A_u^{(n-1)}-A_r^{(n-1)}\big)+D \cdot A^{(n-2)}\,,\\
		\label{eq:A_u}&(2-2n)\pr_u A_u^{(n)}=[\Delta + 	(n-1)(n-2)]A_u^{(n-1)}\,,\\	
		\label{eq:A_r}&(2-2n)\pr_u A^{(n)}_r = 	[\Delta+n(n-3)]A_r^{(n-1)} + 2\, A_u^{(n-1)}-2 D\cdot A^{(n-2)}\,,\\
		\label{eq:A_i}&-2n\,\pr_u A_i^{(n)}=[\Delta+n(n-1)-1]A_i^{(n-1)}-2D_i(A_u^{(n)}-A_r^{(n)})\,.
	\end{align}
\end{subequations}
The last equation can be split by means of \eqref{split_A} and we find
\begin{subequations}
	\begin{align}
		\label{eq:A}&-2n\,\pr_u A^{(n)}=[\Delta+n(n-1)]A^{(n-1)}-2(A_u^{(n)}-A_r^{(n)})\,\\
		\label{eq:A'}&-2n\,\pr_u A^{\p\,(n)}=[\Delta+n(n-1)]A_i^{(\p\,n-1)}\,.
	\end{align}
\end{subequations}
where we used that $ (\D-1) D_iA=D_i\D A $. 
\subsection{Two-form}
The wave equation is
\begin{subequations}
	\begin{align}
		\Box \cB_{ur} &= \left(\pr_r^2 -2\pr_u\pr_r +\frac{\D}{r^2}  -\frac{2}{r}(\pr_u -\pr_r) -\frac{2}{r^2}\right) \cB_{ur} -\frac{2}{r^3}D^i\cB_{ui}\,,\\
		\Box \cB_{ui}&=\left(\pr_r^2 -2\pr_u\pr_r +\frac{\D}{r^2}   -\frac{1}{r^2}\right)\cB_{ui} + \frac{2}{r}D_i\cB_{ur}\,,\\
		\Box \cB_{ri} &= \left(\pr_r^2 -2\pr_u\pr_r +\frac{\D}{r^2}   -\frac{1}{r^2}\right) \cB_{ri} +\frac{2}{r^3}D^j \cB_{ij}+\frac{2}{r}D_i\cB_{ur}\,,\\
		\Box \cB_{ij} &= \left(\pr_r^2 -2\pr_u\pr_r +\frac{\D}{r^2}   +\frac{2}{r}(\pr_u-\pr_r)\right)\cB_{ij} -\frac{2}{r}\big(D_i(\cB_{uj}-\cB_{rj})  -D_j(\cB_{ui}-\cB_{ri})\big)\,,
	\end{align}
\end{subequations}
while the Lorenz gauge condition $ G_\n :=\nabla^\m \cB_{\m\n} = 0 $ implies
\begin{subequations}
	\begin{align}
		&G_u = \left(\pr_u-\pr_r -\frac{2}{r}\right)\cB_{ur}-\frac{1}{r^2}D^i \cB_{ui}\,,\\
		&G_r=-\left(\pr_r +\frac{2}{r}\right)\cB_{ur}-\frac{1}{r^2}D^i\cB_{ri}\,,\\
		&G_i = (\pr_r -\pr_u)\cB_{ri}-\pr_r \cB_{ui}-\frac{1}{r^2}D^j\cB_{ij}\,.
	\end{align}
\end{subequations}

In our setup the logarithms appear only in pure gauge sectors and therefore trivially solve the equations of motion, implying non-trivial conditions only in the $ B- $part of the $ \cB $ field. However, here we analyse the order-by-order equations for a two form gauge field polyhomogeneously expanded.

In particular, the order-by-order wave equation for the logarithmic part is\footnote{In this case, moving to the right the l.h.s.\ of each equation we recover the order-by-order expression of $ \left[ \Box \cB_{\m\n} \right]^{(n)} $.}
\begin{subequations}	
	\begin{align}
		-2(n-2)\pr_u {\hat\cB}_{ur}^{(n-1)}=&(\D + (n-1)(n-4)){\hat\cB}_{ur}^{(n-2)}-2D^i{\hat\cB}_{ui}^{(n-3)}\\
		-2(n-1)\pr_u {\hat\cB}_{ui}^{(n-1)}=&(\D + (n-1)(n-2)-1){\hat\cB}_{ui}^{(n-2)}+2D_i{\hat\cB}_{ur}^{(n-1)}\\
		-2(n-1)\pr_u {\hat\cB}_{ri}^{(n-1)}=&(\D + (n-1)(n-2)-1){\hat\cB}_{ri}^{(n-2)}+2D_i{\hat\cB}_{ur}^{(n-1)}+2D^j{\hat\cB}_{ij}^{(n-3)}\\
		- 2n \pr_u {\hat\cB}_{ij}^{(n-1)} =& (\D + (n+1)(n-2)){\hat\cB}_{ij}^{(n-2)}-2D_i({\hat\cB}_{uj}^{(n-1)}-{\hat\cB}_{rj}^{(n-1)})\notag\\&+2D_j({\hat\cB}_{ui}^{(n-1)}-{\hat\cB}_{ri}^{(n-1)})
	\end{align}
\end{subequations}
while for the rest we find
\begin{subequations}	
	\begin{align}
		-2(n-2)\pr_u \cB_{ur}^{(n-1)}+2\pr_u{\hat\cB}_{ur}^{(n-1)}=&(\D + (n-1)(n-4))\cB_{ur}^{(n-2)}-(2n-5){\hat\cB}_{ur}^{(n-2)}\notag\\ &-2D^i\cB_{ui}^{(n-3)}\\
		-2(n-1)\pr_u \cB_{ui}^{(n-1)}+2\pr_u {\hat\cB}_{ui}^{(n-1)}=&(\D + (n-1)(n-2)-1)\cB_{ui}^{(n-2)}-(2n-3){\hat\cB}_{ui}^{(n-2)}\notag \\ &+2D_i\cB_{ur}^{(n-1)}\\
		-2(n-1)\pr_u \cB_{ri}^{(n-1)}+2\pr_u {\hat\cB}_{ri}^{(n-1)}=&(\D + (n-1)(n-2)-1)\cB_{ri}^{(n-2)}\notag \\ &-(2n-3){\hat\cB}_{ri}^{(n-2)}+2D_i\cB_{ur}^{(n-1)}+2D^j\cB_{ij}^{(n-3)}\\
		- 2n \pr_u \cB_{ij}^{(n-1)}+2\pr_u {\hat\cB}_{ij}^{(n-1)} =& (\D + (n+1)(n-2))\cB_{ij}^{(n-2)}-(2n-1){\hat\cB}_{ij}^{(n-2)}\notag\\&-2D_i(\cB_{uj}^{(n-1)}-\cB_{rj}^{(n-1)})+2D_j(\cB_{ui}^{(n-1)}-\cB_{ri}^{(n-1)}).
	\end{align}
\end{subequations}

The order-by-order gauge condition for the logarithms is
\begin{subequations}	
	\begin{align}
		&-\pr_u {\hat\cB}_{ur}^{(n)} = (n-3){\hat\cB}_{ur}^{(n-1)} - D^i{\hat\cB}_{ui}^{(n-2)} \\		
		& (n-3){\hat\cB}_{ur}^{(n-1)} - D^i{\hat\cB}_{ri}^{(n-2)}=0\\	
		&\pr_u {\hat\cB}_{ri}^{(n)}=(n-1)({\hat\cB}_{ui}^{(n-1)}- {\hat\cB}_{ri}^{(n-1)})-D^j{\hat\cB}_{ij}^{(n-2)}
	\end{align}
\end{subequations}
while
\begin{subequations}	
	\begin{align}
		&-\pr_u \cB_{ur}^{(n)}= -{\hat\cB}_{ur}^{(n-1)} + (n-3)\cB_{ur}^{(n-1)} - D^i\cB_{ui}^{(n-2)} \\		
		& (n-3)\cB_{ur}^{(n-1)} -{\hat\cB}_{ur}^{(n-1)} - D^i\cB_{ri}^{(n-2)}=0 \\		
		&\pr_u \cB_{ri}^{(n)}={\hat\cB}_{ri}^{(n-1)}-{\hat\cB}_{ui}^{(n-1)} +(n-1)(\cB_{ui}^{(n-1)} -\cB_{ri}^{(n-1)})-D^j\cB_{ij}^{(n-2)}
	\end{align}
\end{subequations}

For completeness, the following are the equations for the part that admits a $ \tfrac{1}{r} $ expansion, which are identical to the previous ones for the logarithmic part. 
The gauge condition is 
\begin{subequations}	
	\begin{align}
		&-\pr_u B_{ur}^{(n)} = (n-3)B_{ur}^{(n-1)} - D^iB_{ui}^{(n-2)} \\		
		& (n-3)B_{ur}^{(n-1)} - D^iB_{ri}^{(n-2)}=0\\	
		&\pr_u B_{ri}^{(n)}=(n-1)(B_{ui}^{(n-1)}- B_{ri}^{(n-1)})-D^jB_{ij}^{(n-2)}
	\end{align}
\end{subequations}
and the equations of motion are
\begin{subequations}	
	\begin{align}
		-2(n-2)\pr_u B_{ur}^{(n-1)}=&(\D + (n-1)(n-4))B_{ur}^{(n-2)}-2D^iB_{ui}^{(n-3)}\\
		-2(n-1)\pr_u B_{ui}^{(n-1)}=&(\D + (n-1)(n-2)-1)B_{ui}^{(n-2)}+2D_iB_{ur}^{(n-1)}\\
		-2(n-1)\pr_u B_{ri}^{(n-1)}=&(\D + (n-1)(n-2)-1)B_{ri}^{(n-2)}+2D_iB_{ur}^{(n-1)}+2D^jB_{ij}^{(n-3)}\\
		- 2n \pr_u B_{ij}^{(n-1)} =& (\D + (n+1)(n-2))B_{ij}^{(n-2)}-2D_i(B_{uj}^{(n-1)}-B_{rj}^{(n-1)})\notag\\&+2D_j(B_{ui}^{(n-1)}-B_{ri}^{(n-1)})
	\end{align}
\end{subequations}

By means of the splitting \eqref{B_r-B_u} and the writing \eqref{Bij}, we can write the equations for $ B_r^{\p}, B_r^{\p} $ and  $ B $, which are relevant for the field strength. In particular, we have
\begin{subequations}
	\begin{align}
		&\pr_u B_{r}^{\p\,(n)}=(n-1)(B_{u}^{\p\,(n-1)}- B_{r}^{\p\,(n-1)})-B^{(n-2)},\\
		&-2(n-1)\pr_u B_{u}^{\p\,(n-1)}=(\D + (n-1)(n-2))B_{u}^{\p\,(n-2)}\\
		&-2(n-1)\pr_u B_{r}^{\p\,(n-1)}=(\D + (n-1)(n-2))B_{r}^{\p\,(n-2)}+2B^{(n-3)}\\
		&- 2n \pr_u B^{(n-1)} = (\D + (n+1)(n-2))B^{(n-2)}
		+4\D(B_{u}^{\p\,(n-1)}-B_{r}^{\p\,(n-1)})
	\end{align}
\end{subequations}

\section{Solutions of the equations of motion}\label{app:eom_sol}
	
\subsection{ $ F_{ur} $.}
Let us start employing the complete gauge fixing mentioned in the text. Using the subleading integration functions we reach a setup with 
\begin{equation}
	\nabla^\m A_\m=0,\qquad 	\Box A_\m=0,\qquad A_u^{(1)}=0,\qquad A_{r}^{(n,0)}=0,
\end{equation}
where $ A_{r}^{(n,0)} $ is the $ u- $independent part of $ A_{r}^{(n)} $ near $ \cI^+_- $. We do not have further free subleading terms in the residual gauge parameter $ \e $. 

First, we solve the order-by-order equation $ \Box A_u $, which reads
	\begin{equation}
		(2-2n)\pr_u A_u^{(n)}=[\Delta + 	(n-1)(n-2)]A_u^{(n-1)}\,,
	\end{equation}
and use $ A_u^{(1)}=0 $.  We start with a $ A_u^{(2)}=  A_{u,c}^{(2)}(z,\bz)$ and the equation gives
\begin{equation}
	A_u^{(n)} = \sum_{m=0}^{n-2} (-u)^m A_u^{(n,m)},
\end{equation}
with 
\begin{equation}
	A_u^{(n,m)} =\frac{1}{2^{m}} \frac{(n-m-1)!}{ m! (n-1)! }\prod_{i=1}^{m} \left[\D + i(i+1)\right]A_{u,c}^{(n-m)}\,.
\end{equation}
This result is exact at $ \cI^+ $. Now we can use the other gauge conditions to determine a relation between $A_{u,c}^{(n-m)} $ and $ A $. 

The equations $ \Box A_r $ and $ \nabla^\m A_\m $ give a order-by-order constraint
\begin{equation}\label{constraint_A}
	[\D - (n-2)(n-3)]A_r^{(n-1)}+2(n-2)^{2}A_u^{(n-1)}+2(n-2)D\cdot A^{(n-2)}=0
\end{equation}
Let us focus on the limit $ u\to-\infty $ and, in particular, to the $ u- $independent part. The condition $ A_{r}^{(n,0)}=0 $ implies that 
\begin{equation}
	A_{u}^{(n,0)}= -\frac{1}{n-1}D^i A_{i}^{(n-1,0)},
\end{equation}
and therefore we are able to find all the $ A_u^{(n)} $ in terms of the free Cauchy data
\begin{equation}\label{Au_A}
	A_u^{(n,m)} =-\frac{1}{2^{m}} \frac{(n-m-2)!}{ m! (n-1)! }\prod_{i=1}^{m} \left[\D + i(i+1)\right]D^iA_{i}^{(n-m,0)}.
\end{equation}

Having determined the expression of $ A_u $ in terms of the free data, the equation \eqref{constraint_A} allow us to determine the $ A_r $ as well. However, we cannot extract an exact form on $ \cI^+ $ because $ A_i $ as an arbitrary dependence on $ u $. 

Still, the gauge condition $ A_r^{(n,0)}=0 $ allow us to determine in a compact form the structure of $ 	F_{ur}^{(n,0)} $. In fact, substituting it in \eqref{fur} and using the Helmholtz decomposition, we find
\begin{equation}\label{app:fur}
	\begin{split}
		F_{ur}^{(2,0)} &= \D A^{(0,0)}\,\\
		F_{ur}^{(n,0)} &= -\D A^{(n-2,0)}\qquad (n> 2)\,.
	\end{split}
\end{equation}

\subsection{ $ H_{uri} $}
Let us now discuss the component $ H_{uri} $, which enters in the two-form asymptotic charge. We have by definition
\begin{equation}
	H_{uri}^{(n)} = \pr_u B_{ri}+(n-1)B_{ui}^{(n-1)}+D_i B_{ur}.
\end{equation}
and substituting the Lorenz gauge condition on $ \pr_u B_{ri} $ we find
\begin{equation}
	H_{uri}^{(n)} = (n-1)\left(2B_{ui}^{(n-1)}-B_{ri}^{(n-1)}\right)+D_i B_{ur}-D^j B_{ij}^{(n-2)}.
\end{equation}
Since $ D^i H_{uri}=0 $, we can write 
\begin{equation}
	H_{uri}^{(n)} = \ve_{ij}D^j H_{ur}^{\p\,(n)}
\end{equation}
with 
\begin{equation}
	H_{ur}^{\p\,(n)} = (n-1)\left(2B_{u}^{\p\,(n-1)}-B_{r}^{\p\,(n-1)}\right)-B^{(n-2)}\,.
\end{equation}
Combining equations of motion with the gauge condition we find 
\begin{equation}\label{constraint_B}
	[\D-n(n-1)]B_{r}^{\p\,(n-1)} + 2n(n-1)B_{u}^{\p\,(n-1)}-2(n-1)B^{(n-2)}=0\,,
\end{equation}
which leads to 
\begin{equation}
	H_{ur}^{\p\,(n)} = -\frac{\D}{n}B_{r}^{\p\,(n-1)}+\frac{n-2}{n}B^{(n-2)}.
\end{equation}
Note that the fact that, for $ n=2 $, this object does not depend on $ B^{(0)} $ is simply due to the fact that we are left with a free function $ \l_i^{(0)}(u,z,\bz) $ that can  be used to fix $ B^{(0)}=0. $ Consistently, we can use the subleading integration functions to impose the condition
\begin{equation}
	B_{r}^{\p\,(n,0)}=0 \qquad (n>2).
\end{equation}
This leaves us with 
\begin{equation}
	\begin{split}
	\qquad	H_{ur}^{\p\,(1,0)} =& -B^{(-1,0)},\\
		H_{ur}^{\p\,(2,0)} =& -\frac{\D}{2}B_r^{\p(1,0)},\\
		H_{ur}^{\p\,(n,0)} =& \frac{n-2}{n}B^{(n-2,0)} \qquad (n>2).
	\end{split}
\end{equation}

\bibliographystyle{utphys}

\begin{thebibliography}{10}
	
	\bibitem{Strominger:2017zoo}
	A.~Strominger, {\em {Lectures on the Infrared Structure of Gravity and Gauge
			Theory}}.
	\newblock Princeton University Press, 3, 2017.
	\newblock \href{http://arxiv.org/abs/1703.05448}{{\tt arXiv:1703.05448
			[hep-th]}}.
	
	\bibitem{He:2014cra}
	T.~He, P.~Mitra, A.~P. Porfyriadis, and A.~Strominger, ``{New Symmetries of
		Massless QED},'' \href{http://dx.doi.org/10.1007/JHEP10(2014)112}{{\em JHEP}
		{\bf 10} (2014)  112}, \href{http://arxiv.org/abs/1407.3789}{{\tt
			arXiv:1407.3789 [hep-th]}}.
	
	\bibitem{Pasterski:2015zua}
	S.~Pasterski, ``{Asymptotic Symmetries and Electromagnetic Memory},''
	\href{http://dx.doi.org/10.1007/JHEP09(2017)154}{{\em JHEP} {\bf 09} (2017)
		154}, \href{http://arxiv.org/abs/1505.00716}{{\tt arXiv:1505.00716
			[hep-th]}}.
	
	\bibitem{Strominger:2013lka}
	A.~Strominger, ``{Asymptotic Symmetries of Yang-Mills Theory},''
	\href{http://dx.doi.org/10.1007/JHEP07(2014)151}{{\em JHEP} {\bf 07} (2014)
		151}, \href{http://arxiv.org/abs/1308.0589}{{\tt arXiv:1308.0589 [hep-th]}}.
	
	\bibitem{He:2015zea}
	T.~He, P.~Mitra, and A.~Strominger, ``{2D Kac-Moody Symmetry of 4D Yang-Mills
		Theory},'' \href{http://dx.doi.org/10.1007/JHEP10(2016)137}{{\em JHEP} {\bf
			10} (2016)  137}, \href{http://arxiv.org/abs/1503.02663}{{\tt
			arXiv:1503.02663 [hep-th]}}.
	
	\bibitem{Campiglia:2018see}
	M.~Campiglia, L.~Freidel, F.~Hopfmueller, and R.~M. Soni, ``{Scalar Asymptotic
		Charges and Dual Large Gauge Transformations},''
	\href{http://dx.doi.org/10.1007/JHEP04(2019)003}{{\em JHEP} {\bf 04} (2019)
		003}, \href{http://arxiv.org/abs/1810.04213}{{\tt arXiv:1810.04213
			[hep-th]}}.
	
	\bibitem{Francia:2018jtb}
	D.~Francia and C.~Heissenberg, ``{Two-Form Asymptotic Symmetries and Scalar
		Soft Theorems},'' \href{http://dx.doi.org/10.1103/PhysRevD.98.105003}{{\em
			Phys. Rev. D} {\bf 98} (2018) no.~10, 105003},
	\href{http://arxiv.org/abs/1810.05634}{{\tt arXiv:1810.05634 [hep-th]}}.
	
	\bibitem{Afshar:2018apx}
	H.~Afshar, E.~Esmaeili, and M.~M. Sheikh-Jabbari, ``{Asymptotic Symmetries in
		$p$-Form Theories},'' \href{http://dx.doi.org/10.1007/JHEP05(2018)042}{{\em
			JHEP} {\bf 05} (2018)  042}, \href{http://arxiv.org/abs/1801.07752}{{\tt
			arXiv:1801.07752 [hep-th]}}.
	
	\bibitem{Campoleoni:2017mbt}
	A.~Campoleoni, D.~Francia, and C.~Heissenberg, ``{On higher-spin
		supertranslations and superrotations},''
	\href{http://dx.doi.org/10.1007/JHEP05(2017)120}{{\em JHEP} {\bf 05} (2017)
		120}, \href{http://arxiv.org/abs/1703.01351}{{\tt arXiv:1703.01351
			[hep-th]}}.
	
	\bibitem{Campoleoni:2018uib}
	A.~Campoleoni, D.~Francia, and C.~Heissenberg, ``{Asymptotic symmetries and
		charges at null infinity: from low to high spins},''
	\href{http://dx.doi.org/10.1051/epjconf/201819106011}{{\em EPJ Web Conf.}
		{\bf 191} (2018)  06011}, \href{http://arxiv.org/abs/1808.01542}{{\tt
			arXiv:1808.01542 [hep-th]}}.
	
	\bibitem{Kapec:2015ena}
	D.~Kapec, M.~Pate, and A.~Strominger, ``{New Symmetries of QED},''
	\href{http://dx.doi.org/10.4310/ATMP.2017.v21.n7.a7}{{\em Adv. Theor. Math.
			Phys.} {\bf 21} (2017)  1769--1785},
	\href{http://arxiv.org/abs/1506.02906}{{\tt arXiv:1506.02906 [hep-th]}}.
	
	\bibitem{He:2019jjk}
	T.~He and P.~Mitra, ``{Asymptotic symmetries and Weinberg\textquoteright{}s
		soft photon theorem in Mink$_{d+2}$},''
	\href{http://dx.doi.org/10.1007/JHEP10(2019)213}{{\em JHEP} {\bf 10} (2019)
		213}, \href{http://arxiv.org/abs/1903.02608}{{\tt arXiv:1903.02608
			[hep-th]}}.
	
	\bibitem{He:2019pll}
	T.~He and P.~Mitra, ``{Asymptotic symmetries in (d + 2)-dimensional gauge
		theories},'' \href{http://dx.doi.org/10.1007/JHEP10(2019)277}{{\em JHEP} {\bf
			10} (2019)  277}, \href{http://arxiv.org/abs/1903.03607}{{\tt
			arXiv:1903.03607 [hep-th]}}.
	
	\bibitem{Henneaux:2019yqq}
	M.~Henneaux and C.~Troessaert, ``{Asymptotic structure of electromagnetism in
		higher spacetime dimensions},''
	\href{http://dx.doi.org/10.1103/PhysRevD.99.125006}{{\em Phys. Rev. D} {\bf
			99} (2019) no.~12, 125006}, \href{http://arxiv.org/abs/1903.04437}{{\tt
			arXiv:1903.04437 [hep-th]}}.
	
	\bibitem{Campoleoni:2019ptc}
	A.~Campoleoni, D.~Francia, and C.~Heissenberg, ``{Electromagnetic and color
		memory in even dimensions},''
	\href{http://dx.doi.org/10.1103/PhysRevD.100.085015}{{\em Phys. Rev. D} {\bf
			100} (2019) no.~8, 085015}, \href{http://arxiv.org/abs/1907.05187}{{\tt
			arXiv:1907.05187 [hep-th]}}.
	
	\bibitem{Campoleoni:2020ejn}
	A.~Campoleoni, D.~Francia, and C.~Heissenberg, ``{On asymptotic symmetries in
		higher dimensions for any spin},''
	\href{http://dx.doi.org/10.1007/JHEP12(2020)129}{{\em JHEP} {\bf 12} (2020)
		129}, \href{http://arxiv.org/abs/2011.04420}{{\tt arXiv:2011.04420
			[hep-th]}}.
	
	\bibitem{Fuentealba:2023huv}
	O.~Fuentealba, ``{Asymptotic $ \mathcal{O} $(r) gauge symmetries and
		gauge-invariant Poincar\'e generators in higher spacetime dimensions},''
	\href{http://dx.doi.org/10.1007/JHEP04(2023)047}{{\em JHEP} {\bf 04} (2023)
		047}, \href{http://arxiv.org/abs/2302.13788}{{\tt arXiv:2302.13788
			[hep-th]}}.
	
	\bibitem{Esmaeili:2019mbw}
	E.~Esmaeili, V.~Hosseinzadeh, and M.~M. Sheikh-Jabbari, ``{Source and response
		soft charges for Maxwell theory on AdS$_{d}$},''
	\href{http://dx.doi.org/10.1007/JHEP12(2019)071}{{\em JHEP} {\bf 12} (2019)
		071}, \href{http://arxiv.org/abs/1908.10385}{{\tt arXiv:1908.10385
			[hep-th]}}.
	
	\bibitem{Esmaeili:2021szb}
	E.~Esmaeili and V.~Hosseinzadeh, ``{p-form surface charges on AdS:
		renormalization and conservation},''
	\href{http://dx.doi.org/10.1007/JHEP11(2021)062}{{\em JHEP} {\bf 11} (2021)
		062}, \href{http://arxiv.org/abs/2107.10282}{{\tt arXiv:2107.10282
			[hep-th]}}.
	
	\bibitem{Campoleoni:2023eqp}
	A.~Campoleoni, A.~Delfante, D.~Francia, and C.~Heissenberg, ``{Renormalization
		of spin-one asymptotic charges in AdS$_{D}$},''
	\href{http://dx.doi.org/10.1007/JHEP12(2023)061}{{\em JHEP} {\bf 12} (2023)
		061}, \href{http://arxiv.org/abs/2308.00476}{{\tt arXiv:2308.00476
			[hep-th]}}.
	
	\bibitem{Barnich:2010eb}
	G.~Barnich and C.~Troessaert, ``{Aspects of the BMS/CFT correspondence},''
	\href{http://dx.doi.org/10.1007/JHEP05(2010)062}{{\em JHEP} {\bf 05} (2010)
		062}, \href{http://arxiv.org/abs/1001.1541}{{\tt arXiv:1001.1541 [hep-th]}}.
	
	\bibitem{Campiglia:2017dpg}
	M.~Campiglia, L.~Coito, and S.~Mizera, ``{Can scalars have asymptotic
		symmetries?},'' \href{http://dx.doi.org/10.1103/PhysRevD.97.046002}{{\em
			Phys. Rev. D} {\bf 97} (2018) no.~4, 046002},
	\href{http://arxiv.org/abs/1703.07885}{{\tt arXiv:1703.07885 [hep-th]}}.
	
	\bibitem{Agriela:2023dnw}
	A.~Agriela and M.~Campiglia, ``{Fermionic asymptotic symmetries in massless
		QED},'' \href{http://dx.doi.org/10.1103/PhysRevD.108.065011}{{\em Phys. Rev.
			D} {\bf 108} (2023) no.~6, 065011},
	\href{http://arxiv.org/abs/2307.11171}{{\tt arXiv:2307.11171 [hep-th]}}.
	
	\bibitem{Manzoni:2024agc}
	F.~Manzoni, ``{Axialgravisolitons at infinite corner},''
	\href{http://dx.doi.org/10.1088/1361-6382/ad61b5}{{\em Class. Quant. Grav.}
		{\bf 41} (2024) no.~17, 177001}, \href{http://arxiv.org/abs/2404.04951}{{\tt
			arXiv:2404.04951 [gr-qc]}}.
	
	\bibitem{Barnich:2009se}
	G.~Barnich and C.~Troessaert, ``{Symmetries of asymptotically flat 4
		dimensional spacetimes at null infinity revisited},''
	\href{http://dx.doi.org/10.1103/PhysRevLett.105.111103}{{\em Phys. Rev.
			Lett.} {\bf 105} (2010)  111103}, \href{http://arxiv.org/abs/0909.2617}{{\tt
			arXiv:0909.2617 [gr-qc]}}.
	
	\bibitem{Barnich:2010ojg}
	G.~Barnich and C.~Troessaert, ``{Supertranslations call for superrotations},''
	\href{http://dx.doi.org/10.22323/1.127.0010}{{\em PoS} {\bf CNCFG2010} (2010)
		010}, \href{http://arxiv.org/abs/1102.4632}{{\tt arXiv:1102.4632 [gr-qc]}}.
	
	\bibitem{Campiglia:2016hvg}
	M.~Campiglia and A.~Laddha, ``{Subleading soft photons and large gauge
		transformations},'' \href{http://dx.doi.org/10.1007/JHEP11(2016)012}{{\em
			JHEP} {\bf 11} (2016)  012}, \href{http://arxiv.org/abs/1605.09677}{{\tt
			arXiv:1605.09677 [hep-th]}}.
	
	\bibitem{Peraza:2023ivy}
	J.~Peraza, ``{Renormalized electric and magnetic charges for O(r$^{n}$) large
		gauge symmetries},'' \href{http://dx.doi.org/10.1007/JHEP01(2024)175}{{\em
			JHEP} {\bf 01} (2024)  175}, \href{http://arxiv.org/abs/2301.05671}{{\tt
			arXiv:2301.05671 [hep-th]}}.
	
	\bibitem{Campiglia:2021oqz}
	M.~Campiglia and J.~Peraza, ``{Charge algebra for non-abelian large gauge	symmetries at O(r)},'' \href{http://dx.doi.org/10.1007/JHEP12(2021)058}{{\em
			JHEP} {\bf 12} (2021)  058}, \href{http://arxiv.org/abs/2111.00973}{{\tt
			arXiv:2111.00973 [hep-th]}}.
		

	\bibitem{Nagy:2024dme}
	S.~Nagy, J.~Peraza and G.~Pizzolo,
	``A General Hierarchy of Charges at Null Infinity via the Todd Polynomials,''
	\href{http://arxiv.org/abs/2405.06629}{{\tt arXiv:2405.06629 [hep-th]}}.
	
	\bibitem{Nagy:2024jua}
	S.~Nagy, J.~Peraza and G.~Pizzolo,
	``Infinite-dimensional hierarchy of recursive extensions for all sub$^n$-leading soft effects in Yang-Mills,''
		\href{http://arxiv.org/abs/2407.13556}{{\tt arXiv:2407.13556 [hep-th]}}.
	
	\bibitem{Campiglia:2016efb}
	M.~Campiglia and A.~Laddha, ``{Sub-subleading soft gravitons and large diffeomorphisms},'' \href{http://dx.doi.org/10.1007/JHEP01(2017)036}{{\em
			JHEP} {\bf 01} (2017)  036}, \href{http://arxiv.org/abs/1608.00685}{{\tt
			arXiv:1608.00685 [gr-qc]}}.
	
	\bibitem{Ferrero:2024eva}
	P.~Ferrero, D.~Francia, C.~Heissenberg, and M.~Romoli, ``{Double-copy
		supertranslations},''
	\href{http://dx.doi.org/10.1103/PhysRevD.110.026009}{{\em Phys. Rev. D} {\bf
			110} (2024) no.~2, 026009}, \href{http://arxiv.org/abs/2402.11595}{{\tt
			arXiv:2402.11595 [hep-th]}}.
	
	\bibitem{Nagy:2022xxs}
	S.~Nagy and J.~Peraza,
	``Radiative phase space extensions at all orders in r for self-dual Yang-Mills and gravity,''
	\href{http://dx.doi.org/10.1007/JHEP02(2023)202}{{\em JHEP} {\bf 02} (2023)  202}, 
	\href{http://arxiv.org/abs/2211.12991}{{\tt arXiv:2211.12991 [hep-th]}}.
	
	\bibitem{Campiglia:2021srh}
	M.~Campiglia and S.~Nagy,
	``A double copy for asymptotic symmetries in the self-dual sector,''
	\href{http://dx.doi.org/10.1007/JHEP03(2021)262}{{\em JHEP} {\bf 03} (2021)  262}, 
	\href{http://arxiv.org/abs/2102.01680}{{\tt arXiv:2102.01680 [hep-th]}}.
	
	\bibitem{Bondi:1962px}
	H.~Bondi, M.~G.~J. van~der Burg, and A.~W.~K. Metzner, ``{Gravitational waves
		in general relativity. 7. Waves from axisymmetric isolated systems},''
	\href{http://dx.doi.org/10.1098/rspa.1962.0161}{{\em Proc. Roy. Soc. Lond. A}
		{\bf 269} (1962)  21--52}.
	
	\bibitem{Sachs_Symmetries}
	R.~Sachs, ``{Asymptotic symmetries in gravitational theory},''
	\href{http://dx.doi.org/10.1103/PhysRev.128.2851}{{\em Phys. Rev.} {\bf 128}
		(1962)  2851--2864}.
	
	\bibitem{Penrose:1962ij}
	R.~Penrose, ``{Asymptotic properties of fields and space-times},''
	\href{http://dx.doi.org/10.1103/PhysRevLett.10.66}{{\em Phys. Rev. Lett.}
		{\bf 10} (1963)  66--68}.
	
	\bibitem{Penrose:1964ge}
	R.~Penrose, ``{Republication of: Conformal treatment of infinity},''\href{http://dx.doi.org/10.1007/s10714-010-1110-5}{{\em Gen. Rel. and Grav.} {\bf 43} (2011) 901--922}.
	
	\bibitem{Penrose:1965am}
	R.~Penrose, ``{Zero rest mass fields including gravitation: Asymptotic
		behavior},'' \href{http://dx.doi.org/10.1098/rspa.1965.0058}{{\em Proc. Roy.
			Soc. Lond. A} {\bf 284} (1965)  159}.
	
	\bibitem{Hirai:2018ijc}
	H.~Hirai and S.~Sugishita, ``{Conservation Laws from Asymptotic Symmetry and
		Subleading Charges in QED},''
	\href{http://dx.doi.org/10.1007/JHEP07(2018)122}{{\em JHEP} {\bf 07} (2018)
		122}, \href{http://arxiv.org/abs/1805.05651}{{\tt arXiv:1805.05651
			[hep-th]}}.
	
	\bibitem{Freidel:2019ohg}
	L.~Freidel, F.~Hopfm\"uller, and A.~Riello, ``{Asymptotic Renormalization in
		Flat Space: Symplectic Potential and Charges of Electromagnetism},''
	\href{http://dx.doi.org/10.1007/JHEP10(2019)126}{{\em JHEP} {\bf 10} (2019)
		126}, \href{http://arxiv.org/abs/1904.04384}{{\tt arXiv:1904.04384
			[hep-th]}}.
	
	\bibitem{GAWEDZKI1972307}
	K.~Gaw\c{e}dzki, ``On the geometrization of the canonical formalism in the
	classical field theory,''
	\href{http://dx.doi.org/https://doi.org/10.1016/0034-4877(72)90014-6}{{\em
			Reports on Mathematical Physics} {\bf 3} (1972) no.~4, 307--326}.
	
	\bibitem{Kijowski:1973gi}
	J.~Kijowski, ``{A finite-dimensional canonical formalism in the classical field
		theory},'' \href{http://dx.doi.org/10.1007/BF01645975}{{\em Commun. Math.
			Phys.} {\bf 30} (1973)  99--128}.
	
	\bibitem{Kijowski:1976ze}
	J.~Kijowski and W.~Szczyrba, ``{A Canonical Structure for Classical Field
		Theories},'' \href{http://dx.doi.org/10.1007/BF01608496}{{\em Commun. Math.
			Phys.} {\bf 46} (1976)  183--206}.
	
	\bibitem{Lee:1990nz}
	J.~Lee and R.~M. Wald, ``{Local symmetries and constraints},''
	\href{http://dx.doi.org/10.1063/1.528801}{{\em J. Math. Phys.} {\bf 31}
		(1990)  725--743}.
	
	\bibitem{Wald:1993nt}
	R.~M. Wald, ``{Black hole entropy is the Noether charge},''
	\href{http://dx.doi.org/10.1103/PhysRevD.48.R3427}{{\em Phys. Rev. D} {\bf
			48} (1993) no.~8, R3427--R3431},
	\href{http://arxiv.org/abs/gr-qc/9307038}{{\tt arXiv:gr-qc/9307038}}.
	
	\bibitem{Wald:1999wa}
	R.~M. Wald and A.~Zoupas, ``{A General definition of 'conserved quantities' in
		general relativity and other theories of gravity},''
	\href{http://dx.doi.org/10.1103/PhysRevD.61.084027}{{\em Phys. Rev. D} {\bf
			61} (2000)  084027}, \href{http://arxiv.org/abs/gr-qc/9911095}{{\tt
			arXiv:gr-qc/9911095}}.
	
	\bibitem{Barnich:2001jy}
	G.~Barnich and F.~Brandt, ``{Covariant theory of asymptotic symmetries,
		conservation laws and central charges},''
	\href{http://dx.doi.org/10.1016/S0550-3213(02)00251-1}{{\em Nucl. Phys. B}
		{\bf 633} (2002)  3--82}, \href{http://arxiv.org/abs/hep-th/0111246}{{\tt
			arXiv:hep-th/0111246}}.
	
	\bibitem{Avery:2015rga}
	S.~G. Avery and B.~U.~W. Schwab, ``{Noether\textquoteright{}s second theorem
		and Ward identities for gauge symmetries},''
	\href{http://dx.doi.org/10.1007/JHEP02(2016)031}{{\em JHEP} {\bf 02} (2016)
		031}, \href{http://arxiv.org/abs/1510.07038}{{\tt arXiv:1510.07038
			[hep-th]}}.
	
	\bibitem{Compere:2018aar}
	G.~Comp\`ere and A.~Fiorucci, ``{Advanced Lectures on General Relativity},''
	\href{http://arxiv.org/abs/1801.07064}{{\tt arXiv:1801.07064 [hep-th]}}.
	
	\bibitem{Ruzziconi:2019pzd}
	R.~Ruzziconi, ``{Asymptotic Symmetries in the Gauge Fixing Approach and the BMS
		Group},'' \href{http://dx.doi.org/10.22323/1.384.0003}{{\em PoS} {\bf
			Modave2019} (2020)  003}, \href{http://arxiv.org/abs/1910.08367}{{\tt
			arXiv:1910.08367 [hep-th]}}.
	
	\bibitem{Ciambelli:2022vot}
	L.~Ciambelli, ``{From Asymptotic Symmetries to the Corner Proposal},''
	\href{http://dx.doi.org/10.22323/1.435.0002}{{\em PoS} {\bf Modave2022}
		(2023)  002}, \href{http://arxiv.org/abs/2212.13644}{{\tt arXiv:2212.13644
			[hep-th]}}.
	
	\bibitem{Freidel:2018fsk}
	L.~Freidel and D.~Pranzetti, ``{Electromagnetic duality and central charge},''
	\href{http://dx.doi.org/10.1103/PhysRevD.98.116008}{{\em Phys. Rev. D} {\bf
			98} (2018) no.~11, 116008}, \href{http://arxiv.org/abs/1806.03161}{{\tt
			arXiv:1806.03161 [hep-th]}}.
	
	\bibitem{Anastasiou:2014qba}
	A.~Anastasiou, L.~Borsten, M.~J. Duff, L.~J. Hughes, and S.~Nagy, ``{Yang-Mills
		origin of gravitational symmetries},''
	\href{http://dx.doi.org/10.1103/PhysRevLett.113.231606}{{\em Phys. Rev.
			Lett.} {\bf 113} (2014) no.~23, 231606},
	\href{http://arxiv.org/abs/1408.4434}{{\tt arXiv:1408.4434 [hep-th]}}.
	
	\bibitem{Campiglia:2017xkp}
	M.~Campiglia and L.~Coito, ``{Asymptotic charges from soft scalars in even
		dimensions},'' \href{http://dx.doi.org/10.1103/PhysRevD.97.066009}{{\em Phys.
			Rev. D} {\bf 97} (2018) no.~6, 066009},
	\href{http://arxiv.org/abs/1711.05773}{{\tt arXiv:1711.05773 [hep-th]}}.
	
	\bibitem{Cristofoli:2021vyo}
	A.~Cristofoli, R.~Gonzo, D.~A. Kosower, and D.~O'Connell, ``{Waveforms from
		amplitudes},'' \href{http://dx.doi.org/10.1103/PhysRevD.106.056007}{{\em
			Phys. Rev. D} {\bf 106} (2022) no.~5, 056007},
	\href{http://arxiv.org/abs/2107.10193}{{\tt arXiv:2107.10193 [hep-th]}}.
	
	\bibitem{Lysov:2014csa}
	V.~Lysov, S.~Pasterski, and A.~Strominger, ``{Low\textquoteright{}s Subleading
		Soft Theorem as a Symmetry of QED},''
	\href{http://dx.doi.org/10.1103/PhysRevLett.113.111601}{{\em Phys. Rev.
			Lett.} {\bf 113} (2014) no.~11, 111601},
	\href{http://arxiv.org/abs/1407.3814}{{\tt arXiv:1407.3814 [hep-th]}}.
	
	\bibitem{Conde:2016csj}
	E.~Conde and P.~Mao, ``{Remarks on asymptotic symmetries and the subleading
		soft photon theorem},''
	\href{http://dx.doi.org/10.1103/PhysRevD.95.021701}{{\em Phys. Rev. D} {\bf
			95} (2017) no.~2, 021701}, \href{http://arxiv.org/abs/1605.09731}{{\tt
			arXiv:1605.09731 [hep-th]}}.
	
	\bibitem{DiVecchia:2015jaq}
	P.~Di~Vecchia, R.~Marotta, M.~Mojaza, and J.~Nohle, ``{New soft theorems for
		the gravity dilaton and the Nambu-Goldstone dilaton at subsubleading
		order},'' \href{http://dx.doi.org/10.1103/PhysRevD.93.085015}{{\em Phys. Rev.
			D} {\bf 93} (2016) no.~8, 085015},
	\href{http://arxiv.org/abs/1512.03316}{{\tt arXiv:1512.03316 [hep-th]}}.
	
	\bibitem{DiVecchia:2015oba}
	P.~Di~Vecchia, R.~Marotta, and M.~Mojaza, ``{Soft theorem for the graviton,
		dilaton and the Kalb-Ramond field in the bosonic string},''
	\href{http://dx.doi.org/10.1007/JHEP05(2015)137}{{\em JHEP} {\bf 05} (2015)
		137}, \href{http://arxiv.org/abs/1502.05258}{{\tt arXiv:1502.05258
			[hep-th]}}.
	
	\bibitem{Alessio:2024onn}
	F.~Alessio, P.~Di~Vecchia, and C.~Heissenberg, ``{Logarithmic soft theorems and
		soft spectra},'' \href{http://arxiv.org/abs/2407.04128}{{\tt arXiv:2407.04128
			[hep-th]}}.
		
		\bibitem{Horn:2022acq}
		B.~Horn,
		``Asymptotic symmetries in Bondi gauge and the sub-subleading soft graviton theorem,''
	\href{doi:10.1088/1361-6382/ad0215}{\em Class. Quant. Grav. \textbf{40} (2023) 23}
		\href{https://arxiv.org/abs/2212.02566}{{\tt
		[arXiv:2212.02566 [hep-th]]}}.
	
\end{thebibliography}

\end{document}